\documentclass[prb,twocolumn]{revtex4-1}

\usepackage{graphicx,bm,amsmath}

\begin{document}

\title{Density-functional theory for systems with noncollinear spin: orbital-dependent exchange-correlation
functionals and their application to the Hubbard dimer}

\author{Carsten A. Ullrich}
\affiliation{Department of Physics and Astronomy, University of Missouri, Columbia, Missouri 65211, USA}

\date{\today }

\begin{abstract}
A new class of orbital-dependent exchange-correlation (xc) potentials for applications in noncollinear spin-density-functional
theory is developed. Starting from the optimized effective potential (OEP) formalism for the exact exchange potential---generalized to the noncollinear case---correlation effects are added via a self-consistent procedure inspired by the Singwi-Tosi-Land-Sj\"olander (STLS) method.
The orbital-dependent xc potentials are applied to the Hubbard dimer in uniform and noncollinear magnetic fields and compared to exact diagonalization
and to the Bethe-ansatz local spin-density approximation. The STLS gives the overall best performance for
total energies, densities and magnetizations, particularly in the weakly to moderately correlated regime.
\end{abstract}


\maketitle

\newcommand{\nuu}{n_{\uparrow \uparrow}}
\newcommand{\nud}{n_{\uparrow \downarrow}}
\newcommand{\ndu}{n_{\downarrow \uparrow}}
\newcommand{\ndd}{n_{\downarrow \downarrow}}

\newcommand{\ua}{\uparrow}
\newcommand{\da}{\downarrow}
\newcommand{\fxc}{f^{\rm xc}}
\newcommand{\hxc}{h^{\rm xc}}
\newcommand{\exc}{e_{\rm xc}^{\rm h}}
\newcommand{\Exc}{E_{\rm xc}}
\newcommand{\bfr}{{\bf r}}
\newcommand{\bfm}{{\bf m}}
\newcommand{\bfB}{{\bf B}}
\newcommand{\bfx}{{\bf x}}
\newcommand{\bfX}{{\bf X}}
\newcommand{\bfY}{{\bf Y}}
\newcommand{\bfv}{{\bf v}}

\newcommand{\uun}{\underline{\underline n}}
\newcommand{\uuv}{\underline{\underline v}}
\newcommand{\uud}{\underline{\underline d}}
\newcommand{\uuT}{\underline{\underline T}}
\newcommand{\uuf}{\underline{\underline f}}
\newcommand{\uub}{\underline{\underline b}}
\newcommand{\uum}{\underline{\underline m}}

\newcommand{\vuu}{v_{\uparrow\uparrow}}
\newcommand{\vud}{v_{\uparrow\downarrow}}
\newcommand{\vdu}{v_{\downarrow\uparrow}}
\newcommand{\vdd}{v_{\downarrow\downarrow}}
\newcommand{\buu}{b_{\uparrow\uparrow}}
\newcommand{\bud}{b_{\uparrow\downarrow}}
\newcommand{\bdu}{b_{\downarrow\uparrow}}
\newcommand{\bdd}{b_{\downarrow\downarrow}}
\newcommand{\Auu}{A_{\uparrow\uparrow}}
\newcommand{\Aud}{A_{\uparrow\downarrow}}
\newcommand{\Adu}{A_{\downarrow\uparrow}}
\newcommand{\Add}{A_{\downarrow\downarrow}}

\newcommand{\uu}{\uparrow\uparrow}
\newcommand{\ud}{\uparrow\downarrow}
\newcommand{\du}{\downarrow\uparrow}
\newcommand{\dd}{\downarrow\downarrow}

\newcommand{\CN}{{\cal N}}
\newcommand{\CNi}{{\cal N}^{-1}}

\section{Introduction} \label{sec:1}

Most applications of density-functional theory (DFT)  \cite{Kohn1999,Medvedev2017} are for systems with a fixed spin quantization
axis.\cite{Jacob2012} The Kohn-Sham single-particle wave functions can then be represented as products of spatial orbitals
and up- or down-spinors, and throughout the system the  magnetization $\bfm(\bfr)$ is collinear with the chosen spin quantization axis.
Examples are conventional ferromagnets and antiferromagnets, or paramagnetic systems in a homogeneous
magnetic field.

In general, the direction of $\bfm(\bfr)$ may vary in space;
spin-up and spin-down are then no longer good global quantum numbers. Noncollinear magnetism occurs frequently in
nature and has been the subject of many studies,  \cite{Kubler1988,Sandratskii1998,Ma2015} for instance
in magnetic metals such as $\gamma$-Fe whose ground state features helical spin-density waves,\cite{Sjostedt2002}
in exchange-frustrated solids such as spin glasses and kagome antiferromagnetic lattices, \cite{Balents2010,Zhou2017} and
in molecular magnets and magnetic clusters.\cite{Yamanaka2000,Yamanaka2001,Freeman2004,Peralta2007}

As always in DFT, the key to success is finding good approximations for the exchange-correlation (xc) functional $E_{\rm xc}$. The most widely used
method is the local spin-density approximation (LSDA), assuming a local reference frame in which the local $z$ axis is the spin quantization axis. \cite{Kubler1988,Sandratskii1998} The xc magnetic field is then given by
\begin{equation}
\mu_B \bfB_{\rm xc}^{\rm LSDA}(\bfr) = \frac{\delta E_{\rm xc}^{\rm LSDA}[n,\bfm]}{\delta \bfm(\bfr)}
= \frac{\partial e_{\rm xc}^{\rm unif}}{\partial m} \frac{n(\bfr) \bfm(\bfr)}{m(\bfr)} \:,
\end{equation}
where $e_{\rm xc}^{\rm unif}$ is the xc energy density of a uniform spin-polarized electron gas, and $\mu_B=e\hbar/2m$ is the Bohr magneton.

By construction,
$\bfB_{\rm xc}^{\rm LSDA}(\bfr)$ is always parallel to $\bfm(\bfr)$; likewise for gradient approximations (GGAs)
in which a rotation to a local spin reference frame is made. However, the exact xc functional will also depend on the
transverse gradients of $\bfm(\bfr)$, which causes local magnetic torques. This local xc
torque is important for the description of spin dynamics, \cite{Capelle2001} and it is missing in the LSDA.

There have been several proposals in the literature for xc functionals that include magnetic torque effects.
\cite{Kleinman1999,Katsnelson2003,Sharma2007,Scalmani2012,Bulik2013,Eich2013a,Eich2013b,Pittalis2017} Eich and Gross used the spin-spiral state
of the electron gas as reference system. \cite{Eich2013a}
Scalmani and Frisch \cite{Scalmani2012} proposed a modification of the GGA which leads to nonvanishing magnetic
torques; however, this approximation imposes the unphysical restriction of treating longitudinal and transverse gradients of $\bfm$
equally. \cite{Eich2013b} The optimized effective potential (OEP) approach by Sharma {\it et al.} is free from such restrictions,
but applies only to the exchange-only limit. \cite{Sharma2007} A concise summary of noncollinear electronic structure theory
was recently given by Goings {\em et al.} \cite{Goings2018}

In this paper, a new class of orbital-dependent xc functionals for noncollinear spins is presented, inspired by the
Singwi-Tosi-Land-Sj\"olander (STLS) approach. \cite{Singwi1968,Singwi1981,GiulianiVignale} The basic idea of STLS is to bring correlation
into many-body response functions
via the fluctuation-dissipation theorem, which leads to a self-consistent construction of the local-field factor.
To date, the STLS method has been predominantly applied to the calculation of total
energies and correlation functions of the homogeneous electron gas, including the spin-polarized case.
\cite{Singwi1968,Singwi1981,GiulianiVignale,Lobo1969,Vashishta1972,Holas1987,Moudgil1995a,Moudgil1995b,Calmels1995,Calmels1997a,Calmels1997b,Kim2001,Tas2004,Dobson2004,Kumar2009,Yoshizawa2009}
Hedayati and Vignale found that STLS outperforms the RPA for the half-filled Hubbard model.\cite{Hedayati1989}
The STLS formalism was generalized to inhomogeneous systems by Dobson and coworkers \cite{Dobson2002,Dobson2009} and successfully
applied to calculate correlation energies in spherical atoms and ions, again outperforming the RPA.\cite{Gould2012}

We will generalize the STLS approach (more precisely, its scalar version, sSTLS) to the case of inhomogeneous systems with noncollinear magnetism.
In the first step, we construct the spin-dependent xc kernel $f^{\rm xc}_{\sigma\sigma',\tau\tau'}(\bfr,\bfr')$,
where $\sigma\sigma'$ and $\tau\tau'$ are pairs of spin indices. We then reverse-engineer the corresponding spin-dependent
xc potential via the Slater approximation to the OEP equation,\cite{Slater1951,Talman1976,Krieger1992,Kummel2008}
and derive an expression for the total energy via coupling-constant integration.

We then apply our sSTLS approach to a very simple benchmark system: the half-filled Hubbard dimer, i.e., two electrons on
two lattice sites with a contact interaction.\cite{Carrascal2015,Carrascal2018,Balcerzak2017,Balcerzak2018} We consider the Hubbard dimer in the presence of
inhomogeneous potentials and both uniform and noncollinear magnetic fields, and calculate the resulting ground-state energies, densities, and magnetizations.
Systematic comparisons are carried out between sSTLS, exact exchange, and LSDA, and
exact results obtained by diagonalization of the full two-body Hamiltonian. We find that the sSTLS gives the best overall performance, in particular
when the interactions are not too strong.

This paper is organized as follows: in Section \ref{sec:2} we give a summary of the DFT formalism for noncollinear spins;
two formally equivalent versions will be presented, based on the spin-density matrix and based on the density-magnetization vector.
Section \ref{sec:3} then discusses orbital-dependent xc potentials for noncollinear spins, in exchange-only and including
correlation using the sSTLS formalism. Section \ref{sec:4} then presents applications to the Hubbard dimer at half filling.
Conclusions and an outlook are given in Section \ref{sec:5}.

\section{DFT for noncollinear magnetic systems: Kohn-Sham formalism} \label{sec:2}

In the following, we will be concerned with $N$-electron systems in the presence of magnetic fields $\bfB(\bfr)$ which
couple to the electronic spins only, giving rise to a magnetization $\bfm(\bfr)$. This includes the case of spontaneous magnetization,
such as in ferromagnetic or antiferromagnetic systems.
Orbital magnetism, which would require current-DFT, \cite{Vignale1987,Vignale1988} will not be considered in this paper.
Atomic units ($\hbar=e=m=4\pi\epsilon_0=1$) will be used unless noted otherwise.
All magnetic fields will be given in units where $\mu_B$ has been absorbed.

\subsection{Densities} \label{sec:2A}

Standard spin-DFT is formulated for systems whose magnetization $\bfm(\bfr)$ has a fixed spatial orientation for all $\bfr$,
usually taken along the $z$-direction.
The Kohn-Sham single-particle wave functions can then be separated into spin-up and spin-down orbitals.

For noncollinear magnetic systems, where the orientation of the vector field $\bfm(\bfr)$ may vary in space, the
spin (with reference to a fixed spin quantization axis) ceases to be a good quantum number. In this case, the single-particle wave
functions that form the basis of the Kohn-Sham system can be expressed in the following two-component column vector form:
\cite{Barth1972,Gunnarsson1976}
\begin{equation} \label{1.1}
\Psi_i({\bf r}) = \left( \begin{array}{c} \psi_{i\uparrow}({\bf r})\\
\psi_{i\downarrow}({\bf r})\end{array} \right) \;,
\end{equation}
that is, the $\Psi_i({\bf r})$ are mixtures of up- and down-spinor wave functions.
The corresponding complex conjugate row vector is denoted as $\Psi_i^\dagger = (\psi_{i\ua}^*,\psi_{i\da}^*)$.

DFT for noncollinear magnetic systems can be formulated in two alternative ways, using the spin-density matrix
or the density-magnetization vector as basic variables. We will now define both quantities and show how they are related.

The Kohn-Sham spin-density matrix $\uun(\bfr)$, a $2\times 2$ Hermitian matrix, is defined as follows:
\begin{eqnarray}\label{1.2}
\uun({\bf r})
&=&
\sum_{i=1}^{N}\Psi_i({\bf r})\Psi_i^\dagger({\bf r})
\nonumber\\
&=& \sum_{i=1}^{N}\left( \begin{array}{cc}
|\psi_{i\uparrow}({\bf r})|^2&
\psi_{i\uparrow}({\bf r})\psi^*_{i\downarrow}({\bf r})\\[2mm]
\psi_{i\downarrow}({\bf r})\psi^*_{i\uparrow}({\bf r})&
|\psi_{i\downarrow}({\bf r})|^2\end{array}\right)
\nonumber\\
&\equiv&
\left( \begin{array}{cc} \nuu(\bfr) & \nud(\bfr)\\ \ndu(\bfr) & \ndd(\bfr) \end{array} \right).
\end{eqnarray}

Alternatively, we can use the total density $n(\bfr)\equiv m_0(\bfr)$ and the magnetization $\bfm(\bfr)$ as basic variables,
where $m_1$, $m_2$, and $m_3$ denote the $x$, $y$, and $z$-components, respectively, of the magnetization vector $\bfm$.
The $m_i$ are real quantities, defined as
\begin{equation} \label{1.3}
m_i(\bfr) = \mbox{tr} \{ \sigma_i \uun(\bfr)\}, \qquad i=0,1,2,3.
\end{equation}
The inverse relation is given by
\begin{equation}\label{1.4}
\uun (\bfr)
= \frac{1}{2}\sum_{j=0}^3 \sigma_j m_j(\bfr) .
\end{equation}
Here, $\sigma_0$ is the $2\times 2$ unit matrix, and $\sigma_1$, $\sigma_2$, and $\sigma_3$ are the Pauli matrices.

The density and the magnetization can be combined into a 4-vector:
\begin{equation}\label{1.5}
\vec m(\bfr) = \left( \begin{array}{c} m_0(\bfr) \\ m_1(\bfr) \\ m_2(\bfr) \\ m_3(\bfr) \end{array} \right)
=
\left( \begin{array}{c} \nuu(\bfr) + \ndd(\bfr) \\ \nud(\bfr)+\ndu(\bfr) \\ i(\nud(\bfr)-\ndu(\bfr))
\\ \nuu(\bfr)-\ndd(\bfr) \end{array} \right).
\end{equation}
The elements of the spin-density matrix can also be arranged in the form of a 4-vector:
\begin{equation}\label{1.6}
\vec n(\bfr) = \mbox{\bf vec}\{\uun(\bfr)\} =
\left( \begin{array}{c} \nuu(\bfr) \\ \ndu(\bfr) \\ \nud(\bfr) \\ \ndd(\bfr) \end{array} \right),
\end{equation}
where $\bf vec$ is the standard vectorization operator, in which a vector is formed by stacking the columns
of a matrix.

The vectors $\vec n(\bfr)$ and $\vec m(\bfr)$ are related via
\begin{equation}\label{1.7}
\vec m (\bfr)
=
\left( \begin{array}{cccc}
1 & 0 & 0 & 1 \\
0 & 1 & 1 & 0 \\
0 & -i & i & 0 \\
1 & 0 & 0 & -1
\end{array} \right)
\left( \begin{array}{c} \nuu(\bfr) \\ \ndu(\bfr) \\ \nud(\bfr) \\ \ndd(\bfr) \end{array} \right)
\equiv
\uuT \: \vec n(\bfr),
\end{equation}
which defines the $4\times 4$  transformation matrix $\uuT$. The inverse transformation is
\begin{equation} \label{1.8}
\vec n(\bfr)
= \frac{1}{2}
\left( \begin{array}{cccc}
1 & 0 & 0 & 1 \\
0 & 1 & i & 0 \\
0 & 1 & -i & 0 \\
1 & 0 & 0 & -1
\end{array} \right)
\left( \begin{array}{c} m_0(\bfr) \\ m_1(\bfr) \\ m_2(\bfr) \\ m_3(\bfr) \end{array} \right)
= \uuT^{-1} \vec m (\bfr)\:.
\end{equation}
Notice that there is a factor $1/2$ included in  $\uuT^{-1}$. Furthermore, we have
$2\uuT^{-1} =\uuT^{\rm h.c.}$, where h.c. denotes the Hermitian conjugate.

\subsection{Kohn-Sham equations and potentials} \label{sec:2B}

The Kohn-Sham equation for noncollinear magnetic systems couples the up- and down-components of $\Psi_i({\bf r})$:
\begin{equation}\label{1.9}
\sum_{\beta = \uparrow\downarrow}
\left[ - \frac{\nabla^2}{2}\,\delta_{\alpha\beta} +
v^{\rm KS}_{\alpha \beta}({\bf r})  \right]
\psi_{i\beta}({\bf r}) = \epsilon_i \psi_{i\alpha}({\bf r})
\end{equation}
(here and in the following, Greek subscripts denote spin indices $\ua$ and $\da$).
The effective Kohn-Sham potential $v^{\rm KS}_{\alpha \beta}({\bf r})$ is a $2\times 2$ matrix,
consisting, as usual, of external, Hartree and xc parts:
\begin{equation}\label{1.10}
v^{\rm KS}_{\alpha \beta}({\bf r})
=
v^{\rm ext}_{\alpha \beta}({\bf r}) + v^{\rm H}_{\alpha \beta}({\bf r}) + v^{\rm xc}_{\alpha \beta}({\bf r}) \:,
\end{equation}
where the Hartree potential is diagonal in the spin components:
\begin{equation}\label{1.11}
v^{\rm H}_{\alpha \beta}({\bf r}) = \delta_{\alpha\beta} \int d\bfr' \frac{n(\bfr')}{|\bfr - \bfr'|} \:.
\end{equation}
The xc potential is defined as the functional derivative
\begin{equation}\label{1.12}
v^{\rm xc}_{\alpha\beta}(\bfr) = \frac{\delta\Exc[\uun\,]}{\delta n_{\beta\alpha}(\bfr)} \:,
\qquad \alpha,\beta = \ua,\da,
\end{equation}
where the xc energy $E_{\rm xc}[\uun\,]$ is a functional of the spin-density matrix.

Alternatively, the Kohn-Sham equation, Eq. (\ref{1.9}), can be written as
\begin{equation}\label{1.13}
\left[-\frac{\nabla^2}{2}\, \sigma_0 + \vec V^{\rm KS}(\bfr) \cdot \vec\sigma \right]\Psi_i(\bfr)
= \epsilon_i \Psi_i(\bfr) \:.
\end{equation}
Here, $\vec V^{\rm KS}(\bfr)$ and $\vec \sigma$ are 4-vectors, defined as
\begin{equation}\label{1.14}
\vec V^{\rm KS}(\bfr) = \left( \begin{array}{c}
V^{\rm KS}_0(\bfr)\\V^{\rm KS}_1(\bfr)\\V^{\rm KS}_2(\bfr)\\V^{\rm KS}_3(\bfr)
\end{array} \right), \qquad
\vec\sigma = \left( \begin{array}{c} \sigma_0 \\ \sigma_1 \\ \sigma_2 \\ \sigma_3 \end{array}\right).
\end{equation}
The first component of the KS 4-potential, $V^{\rm KS}_0$, corresponds to a scalar potential, and
the other three components, $V^{\rm KS}_{1,2,3}$, form a KS magnetic field, $\bfB^{\rm KS}$.
As before, $\vec V^{\rm KS}(\bfr)$ consists of external, Hartree and xc parts. The Hartree part is purely scalar,
$V_i^{\rm H} \delta_{i0}$.
The xc 4-potential is defined as
\begin{equation} \label{1.15}
V^{\rm xc}_i(\bfr) = \frac{\delta \Exc[\vec m]}{\delta m_i(\bfr)} \:, \qquad i=0,1,2,3.
\end{equation}
Here, the xc energy is written as a functional of $\vec m$, which follows from $E^{\rm xc}[\uun\,]$ via the
transformation (\ref{1.8}).

The connection between the two Kohn-Sham formalisms, Eqs. (\ref{1.9}) and (\ref{1.13}), can be made by
relating
\begin{equation} \label{1.16}
\uuv^{\rm KS}(\bfr) = \sum_{j=0}^3 \sigma_j V^{\rm KS}_j(\bfr) \:,
\end{equation}
or, in 4-vector form, letting $\vec v^{\rm \hspace{0.75mm}KS} = \mbox{\bf vec}\{\uuv^{\rm KS}\}$,
\begin{equation}\label{1.17}
\vec v^{\rm \hspace{0.75mm}KS}(\bfr) = 2 \uuT^{-1} \vec V^{\rm KS}(\bfr) \:.
\end{equation}
Conversely, we have
\begin{equation}\label{1.18}
\vec V^{\rm KS}(\bfr) = \frac{1}{2}\:\uuT \, \vec v^{\rm \hspace{0.75mm}KS}(\bfr) \:.
\end{equation}
It is not difficult to show that
these relations are consistent with definitions (\ref{1.12}) and (\ref{1.15}) of the xc potentials as functional derivatives of the xc energy.


\section{Orbital-dependent xc potentials} \label{sec:3}

\subsection{OEP, KLI, and Slater} \label{sec:3A}

\subsubsection{General formalism}

If the xc energy is given as an explicit functional of $\uun$ or $\vec m$, then the xc potentials follow via straightforward
functional differentiation, see Eqs. (\ref{1.12}) or (\ref{1.15}). However, it is often the case that the xc energy is given
in terms of the Kohn-Sham orbitals, $E_{\rm xc}[\{\psi_{i\sigma}\}]$, which
makes it an implicit density functional. In that case, obtaining the xc potentials requires a different approach,
known as the optimized effective potential (OEP) method.\cite{Talman1976,Krieger1992,Kummel2008}

In Section I of the Supplemental Material\cite{supp} we will provide the details of the derivation of the OEP equation,
starting from the definition (\ref{1.12}) of the xc potential.
The full OEP equation is as follows:\cite{Sharma2007}
\begin{eqnarray} \label{oep}
0 &=&
\sum_i^{N} \sum_{j\ne i}^\infty \int d\bfr'
\left(\sum_{\alpha\beta}v^{\rm xc}_{\alpha \beta}(\bfr')\psi_{i\beta}(\bfr')
-\sum_\alpha \frac{\delta E_{\rm xc}}{\delta \psi^*_{i\alpha}(\bfr')}
\right)
\nonumber\\
&&
\times
\frac{ \psi_{i\mu}^*(\bfr) \psi_{j\nu}(\bfr) \psi_{j\alpha}^*(\bfr')}{\epsilon_j - \epsilon_i}
\; + \; {\rm h.c.}
\end{eqnarray}
This integral equation for the xc potential is formally exact, but in practice it is computationally expensive.
Fortunately, there are several ways in which the OEP can be approximated and significantly simplified.

The so-called KLI approximation \cite{Krieger1992} is obtained by replacing the energy denominators $\epsilon_j - \epsilon_i$ with some
average $\Delta \epsilon$. The OEP integral equation then becomes, after a brief derivation (see Supplemental Material \cite{supp}),
\begin{equation}\label{kli}
\uuv^{\rm xcK} \uun +  \uun\: \uuv^{\rm xcK} =\uub + \uub' \:,
\end{equation}
where xcK stands for the KLI xc potential,
\begin{equation}
b_{\nu\mu}(\bfr) = \sum_i^{N}\left(\frac{\delta E_{\rm xc}}{\delta \psi^*_{i\nu}(\bfr)}\psi_{i\mu}^*(\bfr) + {\rm h.c.}
\right),
\end{equation}
and
\begin{eqnarray}
b'_{\nu\mu}(\bfr) &=&
\sum_i^{N} \psi_{i\mu}^*(\bfr) \psi_{i\nu}(\bfr)\nonumber\\
&\times& \bigg[2\sum_{\alpha\beta} \int d\bfr'
v^{\rm xcK}_{\alpha \beta}(\bfr')\psi_{i\beta}(\bfr')\psi_{i\alpha}^*(\bfr')
\nonumber \\
&-&
\sum_\alpha \int d\bfr' \!
\left(\frac{\delta E_{\rm xc}}{\delta \psi^*_{i\alpha}(\bfr')}\psi_{i\alpha}^*(\bfr') + {\rm h.c.}\right)\bigg].
\end{eqnarray}
Solving the KLI equation, Eq. (\ref{kli}), is much easier than solving the full OEP equation (\ref{oep}) and
can, in fact, be done explicitly.\cite{Krieger1992}

To get the Slater potential (named after Slater's localized Hartree-Fock approximation\cite{Slater1951,Sharp1953}) from the KLI approximation, we discard $\uub'$ in Eq. (\ref{kli}). This gives the following matrix equation:
\begin{equation}\label{Sylvester}
\uuv^{\rm xcS} \uun +  \uun\: \uuv^{\rm xcS} =\uub \:,
\end{equation}
where xcS stands for the Slater xc potential. Equation (\ref{Sylvester})
has the form of a so-called Sylvester equation. Its solution can be expressed
in the following vectorized form:
\begin{equation}\label{Slaterpot}
{\bf vec}\{\uuv^{\rm xcS}\}
=\CNi \: {\bf vec}\{\uub\: \}  \:.
\end{equation}
Here, the $4\times 4$ matrix $\CN$ is defined as
\begin{equation}
\CN = \sigma_0 \otimes \uun + \uun^T \otimes \sigma_0 \:,
\end{equation}
where $\otimes$ denotes the Kronecker matrix product, and $T$ indicates the transpose.

\subsubsection{Exchange-only limit}

The OEP method is most commonly used in the case of exact exchange. For noncollinear magnetic systems,
the exchange energy is
\begin{equation}\label{Ex}
E_{\rm x} =
-\frac{1}{2} \sum_{\sigma\tau} \int\!\!\int \frac{d\bfr d\bfr'}{|\bfr - \bfr'|}
\gamma_{\sigma\tau}(\bfr,\bfr')\gamma_{\tau\sigma}(\bfr',\bfr),
\end{equation}
where the spin-resolved reduced 1-particle Kohn-Sham density matrix is
\begin{equation}
\gamma_{\sigma\tau}(\bfr,\bfr') = \sum_j^N \psi_{j\sigma}(\bfr)\psi_{j\tau}^*(\bfr').
\end{equation}
The elements of the matrix $\uub$ in Eq. (\ref{Sylvester}) then become
\begin{equation} \label{bmat}
b_{\nu\mu}=
-2\sum_\tau \int\frac{ d\bfr'}{|\bfr - \bfr'|}\gamma_{\nu\tau}(\bfr,\bfr') \gamma_{\tau\mu}(\bfr',\bfr)\:.
\end{equation}
In the case of collinear magnetism, with quantization axis along $z$, all transverse ($\ua\da$ and $\da\ua$) terms vanish and
$\uuv^{\rm xc}$ is diagonal.
In the exchange-only case, using Eq. (\ref{bmat}),  we then get the well-known expression\cite{Krieger1992,GiulianiVignale}
\begin{equation}
v_{\sigma\sigma}^{\rm xS}(\bfr) = -\int \frac{d\bfr'}{|\bfr-\bfr'|}
\frac{|\gamma_{\sigma\sigma}(\bfr,\bfr')|^2}{n_{\sigma\sigma}(\bfr)} \:.
\end{equation}

\subsection{STLS for noncollinear spins} \label{sec:3B}

Extending the OEP, KLI, and Slater approaches to include correlation is, in general, quite complicated and
requires a correlation energy functional $E_{\rm c}$ that pairs well with the exact exchange energy functional $E_{\rm x}$.
The book by Engel and Dreizler gives a comprehensive overview of the correlation functionals that have been used in OEP, with various degrees of success.\cite{EngelDreizler}

Here, we pursue a different strategy to include correlation effects. In this Section we present a
generalization of the inhomogeneous STLS approach \cite{Singwi1968,GiulianiVignale,Dobson2002,Dobson2009} to noncollinear spins.
The idea is to first construct an xc kernel from linear-response theory, and then use this kernel to construct (or, perhaps more accurately, to reverse-engineer) an xc potential, using an analogy with the exchange-only Slater potential.

\subsubsection{Noncollinear PGG/Slater exchange kernel} \label{sec:3B1}

Let us begin by considering the  xc kernel of static DFT linear-response theory for noncollinear spin systems, which is defined as
\begin{equation}\label{48}
f^{\rm xc}_{\alpha\beta,\sigma\tau} (\bfr,\bfr')= \frac{\delta^2 E_{\rm xc}}{\delta n_{\beta\alpha}(\bfr)\delta n_{\sigma\tau}(\bfr')} \:.
\end{equation}
Instead of starting with an approximation for $E_{\rm xc}$ and directly evaluating Eq. (\ref{48}), we
follow Petersilka, Gossmann and Gross (PGG)\cite{Petersilka1996,Petersilka1998} and construct an approximate xc kernel
analogous to the definition of the Slater  xc potential, see Eq. (\ref{Sylvester}):
\begin{eqnarray}\label{PGG0}
\lefteqn{\left[\uuf^{\rm xcS} \uun\right]_{\sigma\sigma',\alpha'\alpha}
+ \left[\uun\: \uuf^{\rm xcS} \right]_{\sigma\sigma',\alpha'\alpha}}\nonumber\\
&=&
\sum_i^N \frac{\delta v^{\rm xc}_{\sigma\sigma'}(\bfr)}{\delta \psi^*_{i\alpha'}(\bfr')} \: \psi_{i\alpha}^*(\bfr')
+
\sum_i^N \frac{\delta v^{\rm xc}_{\sigma\sigma'}(\bfr)}{\delta \psi_{i\alpha}(\bfr')} \: \psi_{i\alpha'}(\bfr') \:.
\end{eqnarray}
The remaining task is to calculate the functional derivatives $\delta v^{\rm xc}_{\sigma\sigma'}(\bfr)/\delta \psi^*_{i\alpha'}(\bfr')$
and $\delta v^{\rm xc}_{\sigma\sigma'}(\bfr)/\delta \psi_{i\alpha}(\bfr')$. We substitute the exchange-only Slater potential $v^{\rm xS}_{\sigma\sigma'}(\bfr)$
and, for simplicity, only include those terms in the functional derivatives that act on $b_{\nu\mu}$, Eq. (\ref{bmat}), and kill the integral over $d\bfr'$.
The details of the derivation are given in Section II of the Supplemental Material.\cite{supp} The result for the noncollinear PGG/Slater exchange kernel is
as follows:
\begin{equation}\label{Slaterkernel}
{\bf vec}^2 \{\uuf^{\rm xS} (\bfr,\bfr')\} = [\CNi(\bfr) \otimes {\CNi}^*(\bfr')]{\bf vec}^2 \{\uud^{\rm x} (\bfr,\bfr')\},
\end{equation}
where the double vectorization ${\bf vec}^2$ acting on a tensor with subscripts $\alpha\beta,\sigma\tau$ implies a
stacking with respect to $\sigma \tau$ followed by a stacking with respect to $\alpha\beta$ [see Eq. (32) of the Supplemental Material\cite{supp}].
On the right-hand side, we have
\begin{equation} \label{d1}
d^{\rm x}_{\sigma\sigma',\alpha\alpha'}(\bfr,\bfr')
= -4\frac{\gamma_{\sigma\alpha}(\bfr,\bfr')\gamma_{\alpha'\sigma'}(\bfr',\bfr)}{|\bfr - \bfr'|} \:.
\end{equation}
In the collinear limit, Eq. (\ref{Slaterkernel}) simplifies considerably, and the only nonvanishing elements are
\begin{eqnarray}\label{PGG1}
f^{\rm xS}_{\sigma\sigma,\sigma\sigma}(\bfr,\bfr') &=& -\frac{|\gamma_{\sigma\sigma}(\bfr,\bfr)|^2}{n_{\sigma\sigma}(\bfr)n_{\sigma\sigma}(\bfr')|\bfr - \bfr'|}
\\
f^{\rm xS}_{\sigma\bar\sigma,\sigma\bar\sigma}(\bfr,\bfr') &=& -4\frac{\gamma_{\sigma\sigma}(\bfr,\bfr')\gamma_{\bar\sigma\bar\sigma}(\bfr',\bfr')}{n(\bfr)n(\bfr')|\bfr - \bfr'|} \:,
\label{PGG2}
\end{eqnarray}
where (\ref{PGG1}) is also known as the PGG kernel. \cite{Petersilka1996,Petersilka1998}

A comparison between the so defined exchange kernel and the Slater exchange potential (see Section IID of the Supplemental Material\cite{supp})
shows that
\begin{equation} \label{xcpot}
v_{\sigma\sigma'}^{\rm xS}(\bfr) = \sum_{\tau\tau'}\int d\bfr' f^{\rm xS}_{\sigma\sigma',\tau\tau'}(\bfr,\bfr') n_{\tau\tau'}(\bfr') \:.
\end{equation}
Thus, in this approximation, knowledge of the exchange kernel immediately produces
an approximation for the exchange potential. However, it is important to remember that neither the Slater potential nor the Slater xc kernel can be written
as functional derivatives.

\begin{figure}
\includegraphics[width=\linewidth]{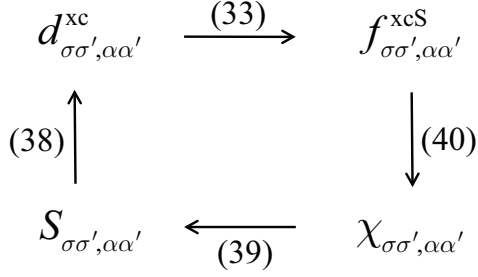}
\caption{Schematic representation of the sSTLS self-consistent construction of the xc kernel. The numbers next to the arrows refer
to equations in the text. } \label{fig1}
\end{figure}

\subsubsection{STLS construction of the xc kernel} \label{sec:3B1a}
We now use the framework derived for the exchange-only case to include correlation. The central idea is to construct
an xc kernel by generalizing the PGG/Slater expression (\ref{Slaterkernel}) and then substitute this into Eq. (\ref{xcpot}) to obtain
an xc potential.

On the right-hand side of Eq. (\ref{Slaterkernel}), instead of $d^{\rm x}$ let us use  the following, more general expression
\begin{eqnarray}\label{d2}
d^{\rm xc}_{\sigma\sigma',\alpha\alpha'}(\bfr,\bfr')
&=&
\frac{4}{|\bfr - \bfr'|} \big[
S_{\sigma\sigma',\alpha\alpha'}(\bfr,\bfr') \nonumber\\
&& - \delta_{\sigma\alpha}\delta(\bfr-\bfr') n_{\alpha'\sigma'}(\bfr)\big] \:,
\end{eqnarray}
where the static structure factor is given by
\begin{equation}\label{S}
S_{\sigma\sigma',\tau\tau'}(\bfr,\bfr') = -\frac{1}{\pi}\int_0^\infty \Im \chi_{\sigma\sigma',\tau \tau'}(\bfr,\bfr',\omega) d\omega,
\end{equation}
involving an integral over the imaginary part of the spin-dependent response function; this follows from the fluctuation-dissipation theorem.\cite{GiulianiVignale}
The spin-density response function of the interacting system can be expressed, using the adiabatic approximation for the xc kernel, as \cite{Ullrich2012}
\begin{eqnarray}\label{fullresponse}
\lefteqn{\chi_{\sigma\sigma',\tau \tau'}(\bfr,\bfr',\omega)
=
\chi^{(0)}_{\sigma\sigma',\tau \tau'}(\bfr,\bfr',\omega)
+
\sum_{\alpha\alpha' \atop \beta\beta'} \int \! d\bfx \int \! d\bfx'} \nonumber\\
&&
\chi^{(0)}_{\sigma\sigma',\alpha\alpha'}(\bfr,\bfx,\omega) f^{\rm Hxc}_{\alpha\alpha',\beta\beta'}(\bfx,\bfx')
\chi_{\beta\beta',\tau \tau'}(\bfx',\bfr',\omega),
\end{eqnarray}
where the Kohn-Sham noninteracting response function is given by \cite{Ullrich2012}
\begin{equation}\label{chi0}
\chi^{(0)}_{\sigma\sigma',\tau \tau'}(\bfr,\bfr',\omega) = \sum_{jk} F_{jk} \frac{\psi_{k\sigma}(\bfr) \psi^*_{j\sigma'}(\bfr)
\psi^*_{k\tau}(\bfr') \psi_{j\tau'}(\bfr')}
{\omega - \varepsilon_k + \varepsilon_j + i\eta}.
\end{equation}
Here, $F_{jk} =f_j-f_k$ is the difference of orbital occupation factors (0 or 1), and $\eta$ is a positive infinitesimal.
In Eq. (\ref{fullresponse}), $f^{\rm Hxc}$ is the sum of the Hartree and xc kernels, where the Hartree kernel is given by
$f^{\rm H}_{\sigma\sigma',\tau\tau'}(\bfr,\bfr') = \delta_{\sigma\sigma'}\delta_{\tau\tau'}/|\bfr-\bfr'|$.

If we evaluate $d^{\rm xc}_{\sigma\sigma',\alpha\alpha'}$ from Eq. (\ref{d2}) with the noninteracting structure factor, which is obtained by using $\chi^{(0)}$ instead of
the full $\chi$ in Eq. (\ref{S}), then (\ref{d2}) reduces to (\ref{d1}) and we recover the Slater exchange kernel $f^{\rm xS}_{\sigma\sigma',\alpha\alpha'}$.

On the other hand, a self-consistent scheme involving Eqs. (\ref{Slaterkernel}), (\ref{d2}), (\ref{S}) and (\ref{fullresponse}),
see Fig. \ref{fig1},
defines a new xc kernel which will improve the Slater exchange kernel by building in correlations.
Substituting this self-consistent xc kernel in Eq. (\ref{xcpot}) then yields an orbital-dependent xc potential.
We call this the scalar STLS (sSTLS) scheme; it simplifies the original STLS approach, \cite{Singwi1968,GiulianiVignale} which, for
inhomogeneous systems, involves xc forces coupling to current densities via a tensorial xc kernel. \cite{Dobson2009}

\subsubsection{Spectral representation of the static structure factor} \label{sec:3B2}

Calculating $S_{\sigma\sigma',\tau\tau'}$ by direct numerical solution of the $\omega$-integral in Eq. (\ref{S})
is inadvisable because of the pole structure of the response function; transforming to the imaginary frequency axis is a better option. \cite{Lein2006} Alternatively, the residues of $\chi_{\sigma\sigma',\tau\tau'}$ can be obtained via a resolvent approach,
\cite{Furche2001} which then makes the $\omega$-integration trivial. We here briefly summarize the results; details are
given in Section III of the Supplemental Material.\cite{supp}

In linear-response time-dependent DFT (TDDFT), the standard method to calculate excitation energies is via the Casida equation.\cite{Ullrich2012,Casida1995}
For the general case of noncollinear spins, the Casida equation reads as follows:
\begin{eqnarray} \label{Casida1}
\lefteqn{\hspace{-3.cm} \sum_{i'a'}\bigg\{ \Big[
\delta_{ii'}\delta_{aa'}  \omega_{a'i'} + K_{ia,i'a'}\Big]X_{i'a'}^{(m)}
+
K_{ia,a'i'} Y_{i'a'}^{(m)} \bigg\}}\nonumber\\
&=& -\Omega_m X_{ia}^{(m)}
\\
\lefteqn{\hspace{-3.cm}  \sum_{i'a'}\bigg\{K_{ai,i'a'}X_{i'a'}^{(m)}
+
\Big[\delta_{aa'}\delta_{ii'} \omega_{a'i'} + K_{ai,a'i'}\Big]Y_{i'a'}^{(m)}\bigg\}}\nonumber\\
&=& \Omega_m Y_{ia}^{(m)}  \:, \label{Casida2}
\end{eqnarray}
where $i,i'$ and $a,a'$ run over occupied and unoccupied orbitals, respectively, and
$\omega_{ai} = \epsilon_a-\epsilon_i$ are the Kohn-Sham excitation energies.
The coupling matrix elements, using the adiabatic Hartree-xc kernel $f^{\rm Hxc}$, are given by
\begin{eqnarray}
K_{pq,p'q'} &=& \sum_{\alpha\alpha'}\sum_{\sigma\sigma'}
\int d\bfx \! \int \! d\bfr \Phi_{p'\alpha q'\alpha'}(\bfx)\nonumber\\
&&\times
 f^{\rm Hxc}_{\alpha\alpha',\sigma\sigma'}(\bfx,\bfr) \Phi^*_{p\sigma q\sigma'}(\bfr) \:,
\end{eqnarray}
where $\Phi_{p\sigma q\sigma'}(\bfr) = \psi^*_{p\sigma}(\bfr) \psi_{q\sigma'}(\bfr)$, and
$p,q,p',q'$ are the orbital indices referred to in Eqs. (\ref{Casida1}) and (\ref{Casida2}).

The Casida equation yields in principle the exact excitation spectrum of the system, provided
the exact xc kernel is used and the exact Kohn-Sham orbitals and energy eigenvalues are given as
input. From the eigenvectors $X_{ia}^{(m)},Y_{ia}^{(m)}$ associated with the $m$th excitation energy $\Omega_m$
one can calculate the transition densities and the oscillator strengths. \cite{Ullrich2012,Casida1995}

In addition, one can use the solutions of the Casida equation to construct a spectral representation of
the interacting response function of the system, Eq. (\ref{fullresponse}). One finds
(see Supplemental Material\cite{supp} for further details)
\begin{eqnarray}
\lefteqn{\chi_{\sigma\sigma',\gamma\gamma'}(\bfr,\bfr',\omega)}\nonumber\\
&=&
\sum_m\sum_{ia \atop i'a'} \frac{\mbox{sign}(\Omega_m)}{\omega - \Omega_m}
  \left[ \Phi^*_{i\sigma a\sigma'}(\bfr)X^{(m)}_{ia}
+ \Phi_{i\sigma' a\sigma}(\bfr) Y^{(m)}_{ia} \right] \nonumber\\
&\times& \left[\Phi_{i'\gamma a' \gamma'}(\bfr')X^{(m)*}_{i'a'}  +   \Phi^*_{i'\gamma' a' \gamma}(\bfr')Y^{(m)*}_{i'a'}  \right].
\end{eqnarray}
This immediately gives the static structure factor:
\begin{eqnarray}
\lefteqn{
S_{\sigma\sigma',\gamma\gamma'}(\bfr,\bfr')}\nonumber\\
&=&
\sum_{m \atop \Omega_m>0}\sum_{ia \atop i'a'}
  \left[ \Phi^*_{i\sigma a\sigma'}(\bfr)X^{(m)}_{ia}
+ \Phi_{i\sigma' a\sigma}(\bfr) Y^{(m)}_{ia} \right] \nonumber\\
&\times& \left[\Phi_{i'\gamma a' \gamma'}(\bfr')X^{(m)*}_{i'a'}  +   \Phi^*_{i'\gamma' a' \gamma}(\bfr')Y^{(m)*}_{i'a'}  \right].
\end{eqnarray}
This expression avoids the frequency integration of Eq. (\ref{S}), but requires solving the Casida equation, which is a routine task for small to moderate system sizes.

\subsubsection{Ground-state energy} \label{sec:3B3}

The total ground-state energy for the noncollinear system described by the Kohn-Sham equation (\ref{1.9}) is
\begin{equation}
E_0 =  \sum_j^{N}\epsilon_j - E_{\rm H}[n]
+ E_{\rm xc}[\uun]
- \int d\bfr \: \mbox{tr}\{\uun(\bfr) \uuv^{\rm xc}(\bfr)\},
\end{equation}
where $E_{\rm H}[n]$ is the Hartree energy. In the sSTLS, we construct an approximate expression for $\uuv^{\rm xc}$,
but not for the xc energy functional $E_{\rm xc}[\uun]$. However, we can obtain $E_{\rm xc}$ using coupling-constant
integration. \cite{Gunnarsson1976} The result, for the noncollinear case, is
\begin{equation}
E_{\rm xc} = \frac{1}{8}\sum_{\alpha\beta} \int d\bfr \int d\bfr' \:
\bar d_{\alpha\alpha,\beta\beta}(\bfr,\bfr') \:,
\end{equation}
where $\bar d$ is obtained from Eq. (\ref{d2}) replacing $S$ by $\bar S$, the coupling-constant averaged structure factor
(see Section IV of the Supplemental Material\cite{supp} for further details).

\section{Asymmetric Hubbard dimer} \label{sec:4}

\subsection{Model} \label{sec:4A}

We now test the performance of our orbital-dependent xc functionals for a simple model system: the Hubbard dimer. \cite{Carrascal2015}
We consider two interacting electrons on two sites, governed by the Hamiltonian
\begin{eqnarray}\label{Hubbard}
\hat H &=& -t\sum_{\sigma} (\hat c^\dagger_{1\sigma}\hat c_{2\sigma} + \hat c^\dagger_{2\sigma}\hat c_{1\sigma}) + U\sum_{l=1,2}
\hat c^\dagger_{l\ua}\hat c_{l\ua}\hat c^\dagger_{l\da}\hat c_{l\da}  \nonumber\\
&+& \sum_{l=1,2}\sum_{\sigma\sigma'}v_{l\sigma\sigma'}^{\rm ext} \hat c^\dagger_{l\sigma}\hat c_{l\sigma'} \:,
\end{eqnarray}
where $\hat c_{l\sigma}^\dagger$ and $\hat c_{l\sigma}$ are creation and annihilation operators, respectively,
for electrons with spin $\sigma$ on site $l$. We fix the hopping parameter as $t=0.5$. The interaction strength $U$
and the on-site external potential $v_{l\sigma\sigma'}^{\rm ext}$ will be varied. The corresponding external
scalar potential and magnetic field are  $V_l^{\rm ext} = \frac{1}{2}\mbox{tr} \, \uuv_{\:l}^{\rm ext}$ and
$\bfB_l^{\rm ext} = \frac{1}{2}\mbox{tr}\, \vec\sigma \uuv_{\: l}^{\rm ext}$, respectively, and similarly for
the Hartree and xc parts.

In the following, we will compare the solutions of the interacting Hamiltonian (\ref{Hubbard}), obtained via exact
diagonalization, with solutions of the corresponding Kohn-Sham system using the exact exchange (EXX) and sSTLS approximations,
as well as the Bethe-ansatz local spin-density approximation (BLSDA). \cite{Franca2012,Capelle2013} For the Hubbard model with
nearest-neighbor interactions, the EXX potential has the particularly simple form
\begin{equation} \label{Hubbardx}
\uuv^{\rm x}_{\:l} = -U \uun_{\:l}
\end{equation}
on each lattice point $l$. This result holds in full OEP as well as in the KLI and Slater approximations for EXX. The sSTLS potential also
becomes simpler, since the xc kernel is local thanks to the Hubbard contact interaction.

\subsection{Parameter space}

In general, the magnetic field can be noncollinear on the two sites. What is the size of the parameter space to be explored?
Nominally, for each $U$, and on each lattice point $l=1,2$, there are 4 parameters for the external potential and magnetic field,
$(V_l^{\rm ext},B_{xl}^{\rm ext},B_{yl}^{\rm ext},B_{zl}^{\rm ext})$, and the corresponding values of density and magnetization are
$(n_l,m_{xl},m_{yl}m_{zl})$. However, the normalization condition $n_1+n_2=2$ and the invariance under a constant potential shift reduce
the dimension by one.

The parameter space can be further reduced by SU(2) spin rotations, which allows one to make a choice of the spin quantization axis.
For instance, one can choose the spin quantization axis to be along the direction of the magnetic field on the second lattice point.
This then leaves us with 4 free magnetic field parameters: $B_{x1}^{\rm ext},B_{y1}^{\rm ext},B_{z1}^{\rm ext}$, and $B_{z2}^{\rm ext}$.

A further reduction of the parameter space comes from an exact property of two-point lattices,\cite{Ullrich2005} namely, for nondegenerate states
the magnetization has the same magnitude on both lattice points, $|\bfm_1| = |\bfm_2|$ (but it may have different directions).
On the other hand, the ground state is invariant under addition of a magnetic field of the form $(\bfB_1,\bfB_2) = \lambda(\bfm_1,-\bfm_2)$, where
$\lambda$ is a constant. This brings down the number of magnetic field parameters to three.

Hence, the space of independent $(V,\bfB)$-parameters of the Hubbard dimer has dimension 4,
plus one additional dimension associated with the interaction strength $U$. In the following, we will not
explore this entire space, but limit ourselves to the collinear case and to a special noncollinear situation.

\subsection{Results} \label{sec:4B}

\begin{figure}[t]
\includegraphics[width=\linewidth]{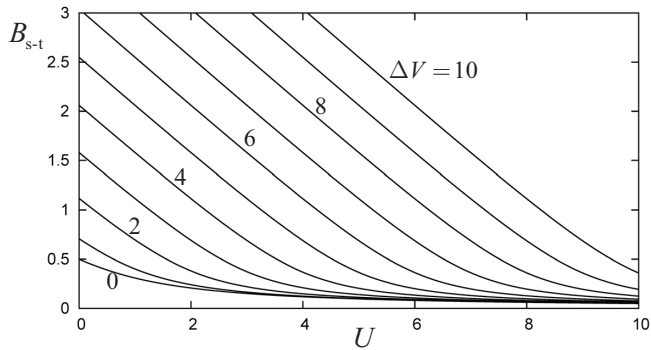}
\caption{Critical magnetic field $B_{\rm s-t}$ at which the half-filled Hubbard dimer makes a transition from singlet to triplet ground state,
for various values of $\Delta V$ between 0 and 10.} \label{fig2}
\end{figure}

\subsubsection{Uniform magnetic field: collinearly restricted approach}

For uniform collinear magnetic fields, exact diagonalization shows that the Hubbard dimer is either in a singlet or in a triplet
ground state, depending on the strength of $\bfB^{\rm ext}$, on $U$, and on $\Delta V = V_1^{\rm ext}-V_2^{\rm ext}$.
Figure \ref{fig2} shows the critical magnetic field $B_{\rm s-t}$
where the singlet-triplet transition occurs, as a function of $U$ and for various values of $\Delta V$.
The magnetic phase diagram is very simple: for magnetic fields up to $B_{\rm s-t}$ the magnetization is zero
on both lattice points. For magnetic fields greater than $B_{\rm s-t}$ the system is fully polarized, with
a magnetization of magnitude 1 and orientation opposite to $\bfB^{\rm ext}$ on both lattice points. In the following,
we will call the former the nonmagnetic (NM) and the latter the ferromagnetic (FM) configuration.

Let us now see what happens in DFT, comparing our three approximations (EXX, sSTLS and BLSDA).
We will consider a Hubbard dimer with scalar potential difference $\Delta V = 2$ and
with a uniform collinear magnetic field along the $z$-direction, $B_{z1}^{\rm ext}=B_{z2}^{\rm ext}=0.2$; other values of $\Delta V$
and $B^{\rm ext}$ lead to similar findings.
From exact diagonalization it follows that the system is NM at small $U$ and then undergoes a transition from
singlet ($m_{z1}=m_{z2}=0$) to triplet ($m_{z1}=m_{z2}=-1$) at $U=3.208$.

\begin{figure}[t]
\includegraphics[width=\linewidth]{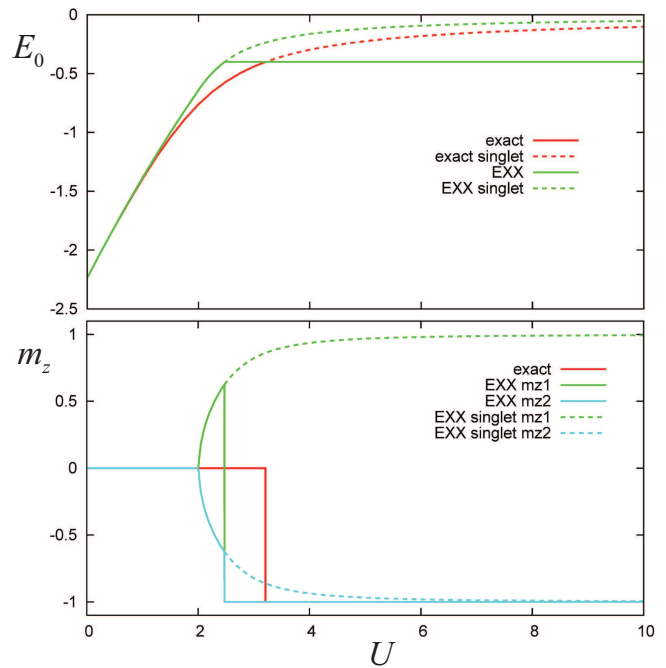}
\caption{(Color online)
Ground-state energy (top) and magnetization (bottom) of the Hubbard dimer with $\Delta V=2$ and $B_{z1}^{\rm ext}=B_{z2}^{\rm ext}=0.2$, comparing collinearly restricted EXX and exact diagonalization. } \label{fig3}
\end{figure}

\begin{figure}[t]
\includegraphics[width=\linewidth]{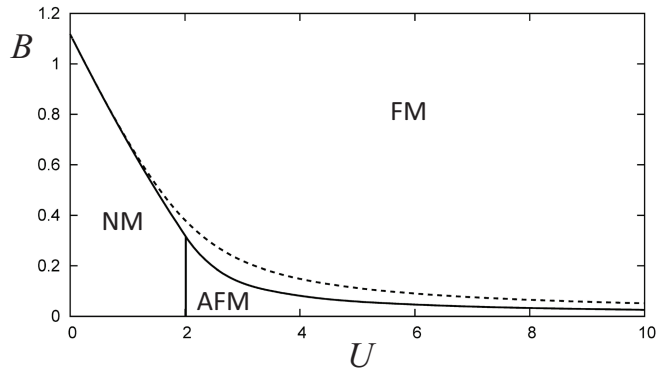}
\caption{Collinearly restricted EXX (full lines) and exact (dashed line) magnetic phase diagrams for the Hubbard dimer with $\Delta V=2$.
In EXX there are three phases: two singlet (NM and AFM) and one triplet (FM).
The exact phase diagram only has the NM and FM phases.} \label{fig4}
\end{figure}

First, we perform EXX Kohn-Sham calculations where the xc magnetic field is restricted to be along the $z$-direction.
The magnetization of the system thus always remains collinear with the applied magnetic field, but, as we will see, is not necessarily
codirectional. The top panel of Fig. \ref{fig3}
compares EXX and exact ground-state energies. We find that the singlet-triplet transition in EXX occurs at $U=2.472$, as indicated by a cusp
in $E_0$ (full green line). There is also a singlet solution of the KS equation for $U>2.472$, indicated by the green dashed line:
it has higher energy than the  FM solution, and approximates the exact singlet ground state (red dashed line).

The bottom panel of Fig. \ref{fig3} shows the associated $z$-components of the magnetization on the two lattice points, which reveals an
interesting feature of  EXX: the system is initially NM for $U<2.013$, then briefly goes through an antiferromagnetic (AFM) phase, for $2.013 < U < 2.472$, before making the transition to the FM phase at $U=2.472$. The symmetry-broken AFM phase, in which $m_{z1}=-m_{z2}$,
turns out to be energetically slightly lower than the NM phase; this, of course, is a manifestation of the well-known ``symmetry dilemma'' of spin-DFT.\cite{Jacob2012,Goings2018}
We here plot the AFM solution where $m_{z1}>0$ and $m_{z2}<0$;
there is an equivalent solution where the signs are reversed.

In Figure \ref{fig4} we compare the exact and the collinearly restricted EXX magnetic phase diagrams for $\Delta V=2$.
As we saw in Fig. \ref{fig2}, the exact solution only has two phases: singlet (NM) and
triplet (FM). By contrast, EXX has two singlet phases, NM and AFM. However, for magnetic field strengths above 0.315 the AFM phase disappears,
since it becomes then energetically favorable to localize the electrons with their magnetic moments aligned.
We also point out that the AFM state at zero magnetic field corresponds to the unrestricted Hartree-Fock solution discussed by Carrascal {\em et al.} \cite{Carrascal2015}

\begin{figure}[t]
\includegraphics[width=\linewidth]{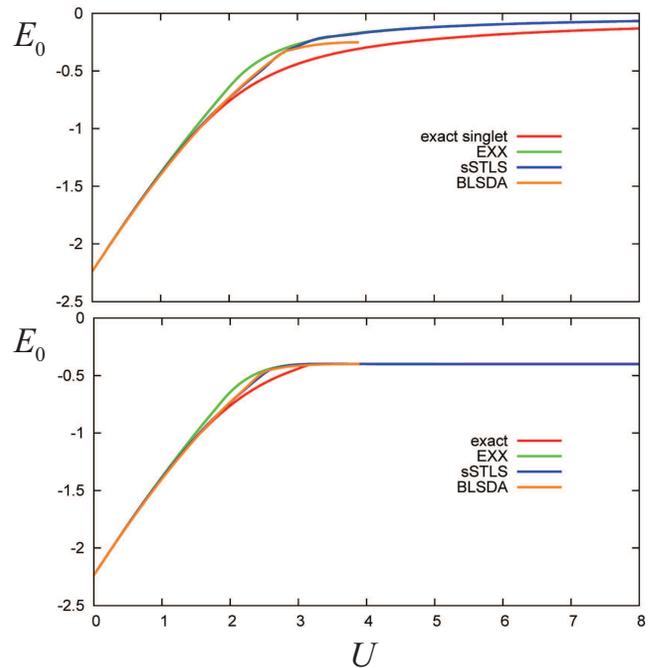}
\caption{(Color online)
Ground-state energy of the Hubbard dimer with $\Delta V=2$ and $B_{z1}^{\rm ext}=B_{z2}^{\rm ext}=0.2$.
Top: collinearly restricted calculation. Bottom: unrestricted calculation, showing the singlet-triplet transition at $U=3.208$. } \label{fig5}
\end{figure}

\begin{figure}[t]
\includegraphics[width=\linewidth]{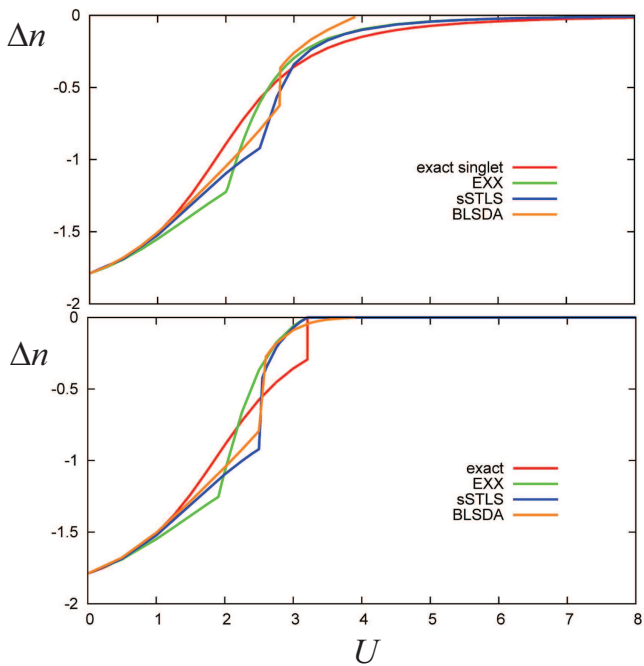}
\caption{(Color online)
Ground-state density difference of the Hubbard dimer with $\Delta V=2$ and $B_{z1}^{\rm ext}=B_{z2}^{\rm ext}=0.2$.
Top: collinearly restricted calculation. Bottom: unrestricted calculation, showing the singlet-triplet transition at $U=3.208$. } \label{fig6}
\end{figure}

Now let us include correlation with BLSDA and sSTLS. It turns out that for the collinearly restricted case these two methods are not applicable to the
triplet (FM) case in which $m_{z1}=m_{z2}=-1$. For the BLSDA, this is well known and comes from the fact
that the  lattice correlation energy density has a discontinuity at half filling;\cite{Franca2012,Capelle2013,Carrascal2015}
therefore, BSLDA Kohn-Sham calculations fail to converge as soon as any lattice site has occupation 1.
In the collinearly restricted sSTLS, on the other hand, spin-flip excitations are not allowed in the Casida equation,
which means that the formalism breaks down as soon as the system is fully spin polarized. However, we will soon see that this
problem goes away in the noncollinear sSTLS formalism.

Thus, in the collinearly restricted case we can only calculate the NM singlet state in BLSDA and the NM and AMF singlet states in sSTLS.
The results are shown in the top part of Fig. \ref{fig5}, which compares the exact singlet ground-state energy $E_0$ with
EXX, sSTLS and BLSDA. The exact singlet $E_0$ starts out linearly and
begins to level off around $U=2$, approaching zero for large interactions. All DFT calculations essentially reproduce this behavior.
BLSDA and sSTLS are extremely close until about $U=3$; after that, sSTLS and EXX are practically on top of each other, whereas the
BLSDA fails to converge for $U>3.95$. The top part of Fig. \ref{fig6} shows the corresponding density differences $\Delta n = n_1 - n_2$,
and we see that for the BLSDA $\Delta n$ approaches zero at $U=3.95$. The differences in $\Delta n$ between the exact result and the DFT results are
more pronounced as for $E_0$. All DFT results have a cusp, indicating the transition from NM to AFM, whereas the exact result is smooth.

\begin{figure}[t]
\includegraphics[width=\linewidth]{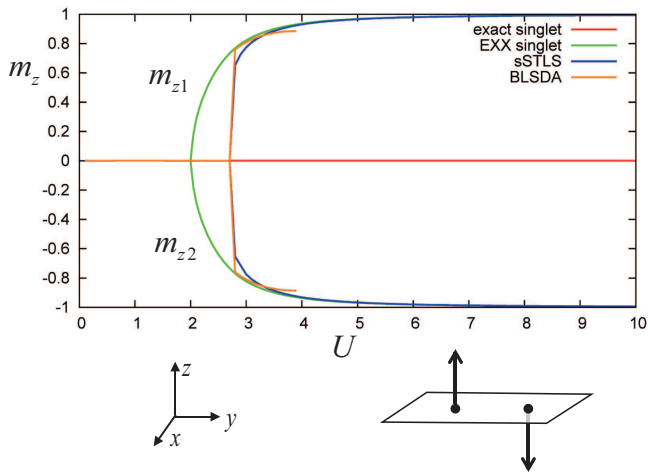}
\caption{(Color online)
Magnetization $m_z$  of the Hubbard dimer with $\Delta V=2$ and $B_{z1}^{\rm ext}=B_{z2}^{\rm ext}=0.2$. The magnetization
is restricted to be along the $z$-direction.   } \label{fig7}
\end{figure}

Let us now look at the magnetization. Figure \ref{fig7} shows the magnetization components $m_{z1}$ and $m_{z2}$ for the collinearly restricted
calculations. The NM to AFM transition occurs at $U=2.013$ in EXX and at $U=2.798$ in BLSDA and sSTLS. After that, the sSTLS quickly merges with the EXX singlet magnetization.

\subsubsection{Uniform magnetic field: unrestricted approach}

We will now consider the same system as above, but lift the restriction of only collinear induced magnetization,
and allow the system to break symmetry to reach the lowest possible
ground-state energy. To do this, we introduce a small transverse magnetic field along the $x$-direction on lattice point 1, and then
let the system find the ground state, which can feature a noncollinear magnetization that is not aligned along the $z$-direction.
The results for $E_0$ are shown in the bottom panel of  Fig. \ref{fig5}. As we can see, the singlet-triplet transition of the
exact system is now reproduced by the Kohn-Sham system, and the agreement is generally very good, except in the crossover region for $U$ between around 2 and 3.

The bottom panel of Fig. \ref{fig6} shows the density difference $\Delta n$. In the exact calculation, $\Delta n$ jumps to zero at the singlet-triplet transition.
The DFT calculations mimic this jump at somewhat smaller values of $U$.

\begin{figure}[t]
\includegraphics[width=\linewidth]{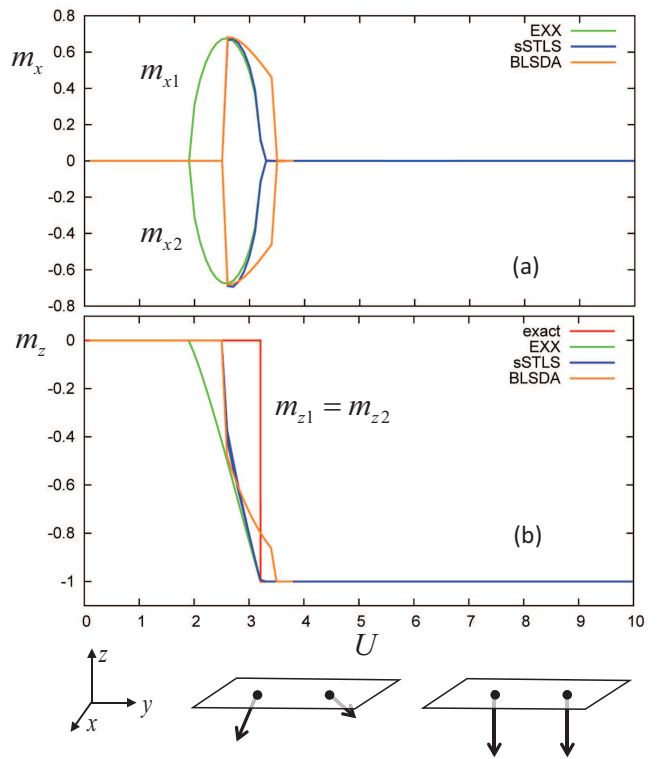}
\caption{(Color online)
Magnetization components (a) $m_x$ and (b) $m_z$ of the Hubbard dimer with $\Delta V=2$ and
$B_{z1}^{\rm ext}=B_{z2}^{\rm ext}=0.2$. Unrestricted
calculation, with symmetry breaking in the crossover region between weak and strong correlations.   } \label{fig8}
\end{figure}

Figure \ref{fig8} shows the magnetization along the $x$ and $z$ directions. The unrestricted DFT calculations all share the same behavior:
instead of making a direct transition from the NM to the FM phase, they pass through a spiral phase which is characterized by
finite values of $m_{x1}$ and $m_{x2}$, with opposite signs. In BLSDA and sSTLS, this spiral phase covers a smaller region of $U$ than
in EXX, and $m_z$ comes closer to the step-like behavior of the exact system. The AFM phase of the collinear restricted calculations
has now disappeared. As before, the BLSDA breaks down as soon as the system is fully localized, i.e., $\Delta n=0$, which happens at $U=3.95$.

We find that the sSTLS is very close to the BLSDA in the weakly correlated regime; as soon as the system begins to approach the crossover
to the localized (Mott) regime, the sSTLS starts to merge with the EXX. Overall, the sSTLS thus shows the best agreement with the exact results.

\subsubsection{Noncollinear magnetic field}

\begin{figure*}[t]
\includegraphics[width=\linewidth]{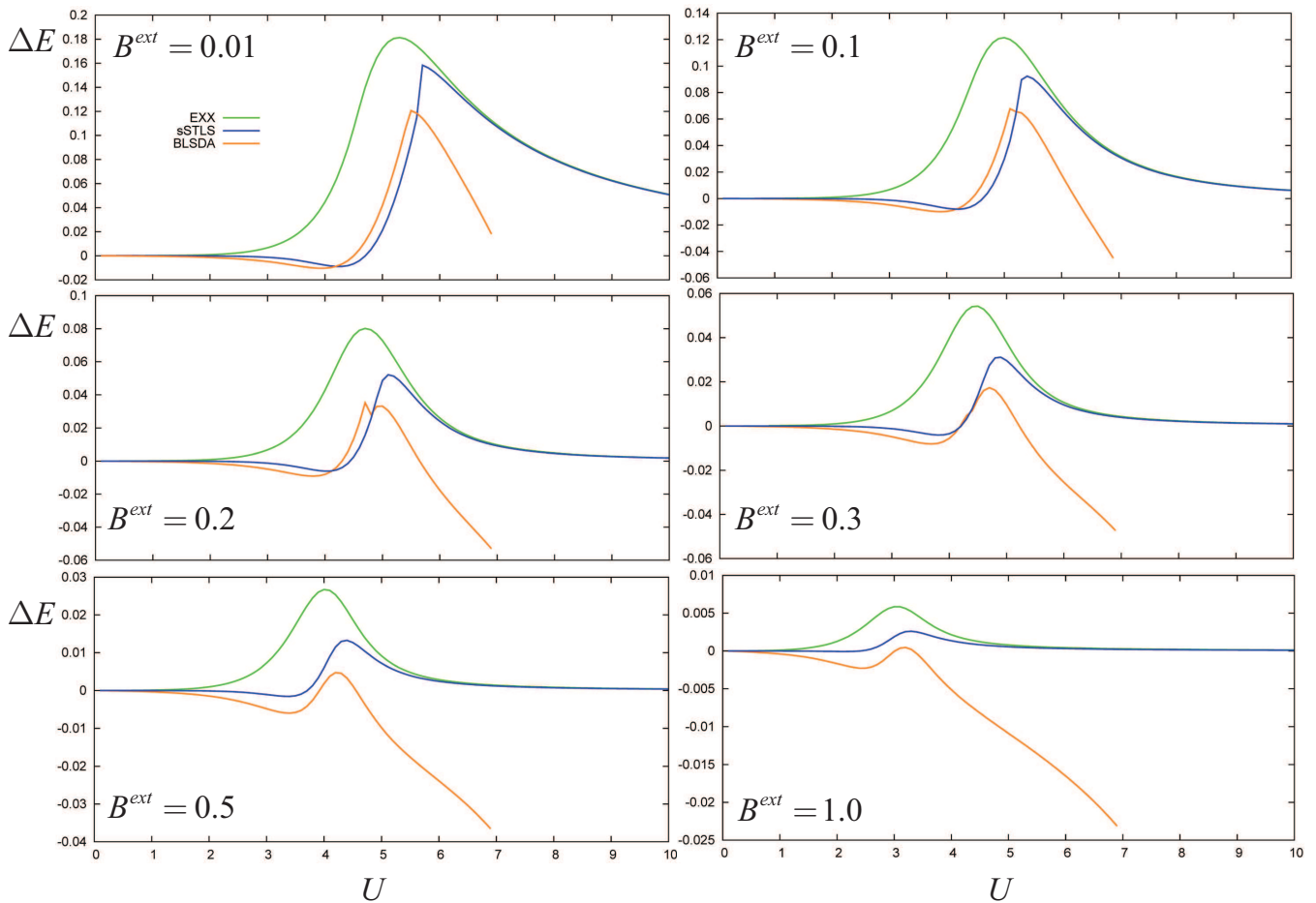}
\caption{(Color online)
Difference between the EXX, sSTLS and BLSDA ground-state energies and the exact ground-state energy for the two-electron Hubbard dimer with $\Delta V=5$.
The applied magnetic field has the same magnitude on both lattice points, as indicated, but is along the $x$-direction on point 1 and along
the $z$-direction on point 2.} \label{fig9}
\end{figure*}

We now consider the case where  the external magnetic field is along the $x$-direction on lattice point 1
and along the $z$-direction on lattice point 2. The magnitudes of $\bfB^{\rm ext}$ are taken to be the same on the two lattice
points, i.e., $B^{\rm ext}_{x1}=B^{\rm ext}_{z2}$. Other situations, for instance different magnitudes of the magnetic field
on the two lattice sites or different angles between the fields on point 1 and 2,  do not lead to essential new physical insight.
The scalar potential difference is taken as $\Delta V = V_1^{\rm ext}-V_2^{\rm ext} = 5$.

In Fig. \ref{fig9} we plot the differences between the EXX, sSTLS and BLSDA ground-state energies and the exact ground-state energy,
$\Delta E = E_0^{\rm approx} - E_0^{\rm exact}$, for values of the magnetic field between 0.01 and 1. Several trends can be observed.

First of all, $\Delta E$ tends to becomes smaller (i.e., the DFT approximations become better) for larger magnetic fields.
The errors are largest in the crossover region where $U$ is comparable
to $\Delta V$; both in the weakly and in the strongly correlated limits the energy errors go to zero, at least for EXX and sSTLS. The BLSDA, on the
other hand, fails in the limit of large correlations, and breaks down at around $U=6.9$. The reason for this, as mentioned earlier, is the
discontinuity of the BLSDA correlation energy density at half filling; the strongly correlated regime associated with the Mott transition causes
the electrons to be localized, so that $n$ approaches 1 on each lattice point. The BLSDA then fails to converge.

The sSTLS clearly gives the best overall results for the total energy, outperforming the BLSDA for weak correlations. For strong correlations ($U \gtrsim 6$) it
coincides with EXX.

\begin{figure*}[t]
\includegraphics[width=\linewidth]{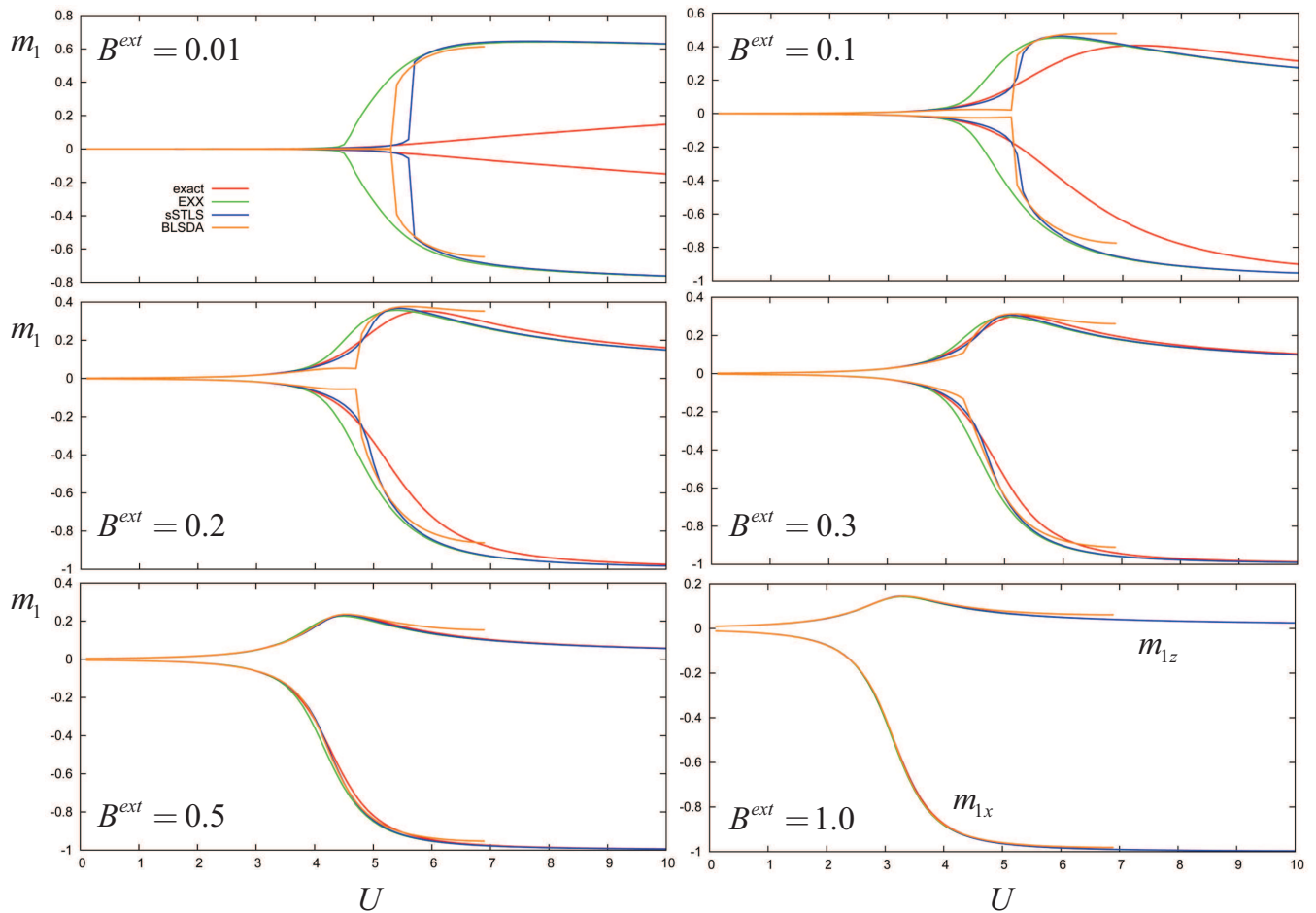}
\caption{(Color online)
Ground-state magnetization components $m_{1x}$ and $m_{1z}$ on lattice point 1 of the two-electron Hubbard dimer with $\Delta V=5$, comparing exact results with
EXX, sSTLS and BLSDA. The applied magnetic field has the same magnitude on both lattice points, as indicated, but is along the $x$-direction on point 1 and along
the $z$-direction on point 2. } \label{fig10}
\end{figure*}

In addition to getting good total energies, it is important for applications in magnetism that the resulting magnetizations are accurate.
In Fig. \ref{fig10} we compare the $x$ and $z$ components of the ground-state magnetization on lattice site 1 following from EXX, sSTLS and BLSDA
with the exact results (on lattice site 2 one finds the same magnetization but with $x$ and $z$ reversed, $m_{2x}=m_{1z}$ and $m_{2z}=m_{1x}$).
Like for the total energies, we find that the agreement for the magnetizations gets better for increasing strength of the applied magnetic fields.

For weak correlations, the sSTLS again performs by far the best. However, in the crossover regime, the error of the magnetization can be
significant with all methods. This is especially the case for $B^{\rm ext}=0.01$: here, the sSTLS is very good until about $U=5.5$, but then jumps to merge with the EXX.
For higher magnetic fields the sSTLS becomes smooth, whereas the BLSDA has cusps at all values of $B^{\rm ext}$ (although less pronounced as the fields get stronger).
These cusps can be viewed
as vestiges of the symmetry-breaking transitions which we have seen to occur in the collinear case discussed above.
In the noncollinear case, at relatively small magnetic fields, these transitions are smoothed out when orbital functionals are used,
but at a local level the BLSDA cannot tell the difference between collinear and noncollinear and hence continues to produce cusps.

\section{Conclusions} \label{sec:5}

In this paper, we have derived and analyzed a new class of orbital-dependent xc functionals, to be used for systems with
noncollinear magnetism. In such systems, the single-particle Kohn-Sham orbitals are mixtures of up and down spinors.
At present, the standard DFT approach is to adapt commonly used xc functionals such as LSDA or GGA to the noncollinear case
by local spin rotations; the underlying reference system is uniformly spin polarized, which leads to unphysical features
in the noncollinear case, most notably, the absence of xc torques. Orbital-dependent xc functionals are free from such
restrictions.

We have presented a detailed derivation of the OEP method for noncollinear spins, including the KLI and Slater approximations.
To build in correlations on top of EXX, we have generalized the STLS approach for inhomogeneous systems, in its simplified scalar version (sSTLS).
To make the sSTLS work in a Kohn-Sham scheme, we construct an xc potential by straightforward integration of the linear-response xc kernel,
in analogy with the exchange-only Slater approximation.

To demonstrate the performance of the sSTLS, we have studied the half-filled Hubbard dimer exposed to collinear and noncollinear magnetic fields.
For the Hubbard model, the BLSDA has been very successfully used to calculate ground-state properties; we here generalize it to the
noncollinear case.

We find that the performance of the BLSDA, EXX and sSTLS depends very strongly on whether the system is weakly or strongly
correlated, which, of course, is not surprising at all. If correlation becomes significant, then a tiny magnetic field can strongly
affect the state of the system, leading to dramatic effects of symmetry breaking. In general, the broken-symmetry solutions
always tend to win: in this way, the system lowers its energy by reducing correlation effects.

We find that, overall, the sSTLS gives the best performance in terms of total energy, density, and magnetization.
At strong correlations, sSTLS approaches the EXX. The BLSDA, on the other hand, suffers a breakdown in the Mott transition region,
but until then its performance is comparable to the sSTLS and close to the exact benchmark results.

The sSTLS thus appears to be a promising method for the study of noncollinear magnetism in matter. However, several important issues
remain to be addressed. On a formal level, one would like to improve the construction of the xc potential from the linear-response xc kernel,
to mitigate the fact that the sSTLS xc potential is not variational. The xc torques obtained from orbital functionals need to be analyzed
and compared with exact benchmarks.
And, finally, orbital functionals for noncollinear magnetism should be implemented and tested for real systems, not just simple models, which
raises questions concerning numerical efficiency.
These issues are currently under investigation and will be reported elsewhere.

\acknowledgments

This work was supported by a Research Corporation for Science Advancement Cottrell Scholar SEED Award.

\bibliography{paper_refs}

\begin{thebibliography}{67}%
\makeatletter
\providecommand \@ifxundefined [1]{%
 \@ifx{#1\undefined}
}%
\providecommand \@ifnum [1]{%
 \ifnum #1\expandafter \@firstoftwo
 \else \expandafter \@secondoftwo
 \fi
}%
\providecommand \@ifx [1]{%
 \ifx #1\expandafter \@firstoftwo
 \else \expandafter \@secondoftwo
 \fi
}%
\providecommand \natexlab [1]{#1}%
\providecommand \enquote  [1]{``#1''}%
\providecommand \bibnamefont  [1]{#1}%
\providecommand \bibfnamefont [1]{#1}%
\providecommand \citenamefont [1]{#1}%
\providecommand \href@noop [0]{\@secondoftwo}%
\providecommand \href [0]{\begingroup \@sanitize@url \@href}%
\providecommand \@href[1]{\@@startlink{#1}\@@href}%
\providecommand \@@href[1]{\endgroup#1\@@endlink}%
\providecommand \@sanitize@url [0]{\catcode `\\12\catcode `\$12\catcode
  `\&12\catcode `\#12\catcode `\^12\catcode `\_12\catcode `\%12\relax}%
\providecommand \@@startlink[1]{}%
\providecommand \@@endlink[0]{}%
\providecommand \url  [0]{\begingroup\@sanitize@url \@url }%
\providecommand \@url [1]{\endgroup\@href {#1}{\urlprefix }}%
\providecommand \urlprefix  [0]{URL }%
\providecommand \Eprint [0]{\href }%
\providecommand \doibase [0]{http://dx.doi.org/}%
\providecommand \selectlanguage [0]{\@gobble}%
\providecommand \bibinfo  [0]{\@secondoftwo}%
\providecommand \bibfield  [0]{\@secondoftwo}%
\providecommand \translation [1]{[#1]}%
\providecommand \BibitemOpen [0]{}%
\providecommand \bibitemStop [0]{}%
\providecommand \bibitemNoStop [0]{.\EOS\space}%
\providecommand \EOS [0]{\spacefactor3000\relax}%
\providecommand \BibitemShut  [1]{\csname bibitem#1\endcsname}%
\let\auto@bib@innerbib\@empty
\bibitem [{\citenamefont {Kohn}(1999)}]{Kohn1999}%
  \BibitemOpen
  \bibfield  {author} {\bibinfo {author} {\bibfnamefont {W.}~\bibnamefont
  {Kohn}},\ }\href@noop {} {\bibfield  {journal} {\bibinfo  {journal} {Rev.
  Mod. Phys.}\ }\textbf {\bibinfo {volume} {71}},\ \bibinfo {pages} {1253}
  (\bibinfo {year} {1999})}\BibitemShut {NoStop}%
\bibitem [{\citenamefont {Medvedev}\ \emph {et~al.}(2017)\citenamefont
  {Medvedev}, \citenamefont {Bushmarinov}, \citenamefont {Sun}, \citenamefont
  {Perdew},\ and\ \citenamefont {Lyssenko}}]{Medvedev2017}%
  \BibitemOpen
  \bibfield  {author} {\bibinfo {author} {\bibfnamefont {M.~G.}\ \bibnamefont
  {Medvedev}}, \bibinfo {author} {\bibfnamefont {I.~S.}\ \bibnamefont
  {Bushmarinov}}, \bibinfo {author} {\bibfnamefont {J.}~\bibnamefont {Sun}},
  \bibinfo {author} {\bibfnamefont {J.~P.}\ \bibnamefont {Perdew}}, \ and\
  \bibinfo {author} {\bibfnamefont {K.~A.}\ \bibnamefont {Lyssenko}},\
  }\href@noop {} {\bibfield  {journal} {\bibinfo  {journal} {Science}\ }\textbf
  {\bibinfo {volume} {355}},\ \bibinfo {pages} {49} (\bibinfo {year}
  {2017})}\BibitemShut {NoStop}%
\bibitem [{\citenamefont {Jacob}\ and\ \citenamefont
  {Reiher}(2012)}]{Jacob2012}%
  \BibitemOpen
  \bibfield  {author} {\bibinfo {author} {\bibfnamefont {C.~R.}\ \bibnamefont
  {Jacob}}\ and\ \bibinfo {author} {\bibfnamefont {M.}~\bibnamefont {Reiher}},\
  }\href@noop {} {\bibfield  {journal} {\bibinfo  {journal} {Int. J. Quantum
  Chem.}\ }\textbf {\bibinfo {volume} {112}},\ \bibinfo {pages} {3661}
  (\bibinfo {year} {2012})}\BibitemShut {NoStop}%
\bibitem [{\citenamefont {K{\"u}bler}\ \emph {et~al.}(1988)\citenamefont
  {K{\"u}bler}, \citenamefont {H{\"o}ck}, \citenamefont {Sticht},\ and\
  \citenamefont {Williams}}]{Kubler1988}%
  \BibitemOpen
  \bibfield  {author} {\bibinfo {author} {\bibfnamefont {J.}~\bibnamefont
  {K{\"u}bler}}, \bibinfo {author} {\bibfnamefont {K.-H.}\ \bibnamefont
  {H{\"o}ck}}, \bibinfo {author} {\bibfnamefont {J.}~\bibnamefont {Sticht}}, \
  and\ \bibinfo {author} {\bibfnamefont {A.~R.}\ \bibnamefont {Williams}},\
  }\href@noop {} {\bibfield  {journal} {\bibinfo  {journal} {J. Phys. F: Met.
  Phys.}\ }\textbf {\bibinfo {volume} {18}},\ \bibinfo {pages} {469} (\bibinfo
  {year} {1988})}\BibitemShut {NoStop}%
\bibitem [{\citenamefont {Sandratskii}(1998)}]{Sandratskii1998}%
  \BibitemOpen
  \bibfield  {author} {\bibinfo {author} {\bibfnamefont {L.~M.}\ \bibnamefont
  {Sandratskii}},\ }\href@noop {} {\bibfield  {journal} {\bibinfo  {journal}
  {Adv. Phys.}\ }\textbf {\bibinfo {volume} {47}},\ \bibinfo {pages} {91}
  (\bibinfo {year} {1998})}\BibitemShut {NoStop}%
\bibitem [{\citenamefont {Ma}\ and\ \citenamefont {Dudarev}(2015)}]{Ma2015}%
  \BibitemOpen
  \bibfield  {author} {\bibinfo {author} {\bibfnamefont {P.-W.}\ \bibnamefont
  {Ma}}\ and\ \bibinfo {author} {\bibfnamefont {S.~L.}\ \bibnamefont
  {Dudarev}},\ }\href@noop {} {\bibfield  {journal} {\bibinfo  {journal} {Phys.
  Rev. B}\ }\textbf {\bibinfo {volume} {91}},\ \bibinfo {pages} {054420}
  (\bibinfo {year} {2015})}\BibitemShut {NoStop}%
\bibitem [{\citenamefont {Sj{\"o}stedt}\ and\ \citenamefont
  {Nordstr{\"o}m}(2002)}]{Sjostedt2002}%
  \BibitemOpen
  \bibfield  {author} {\bibinfo {author} {\bibfnamefont {E.}~\bibnamefont
  {Sj{\"o}stedt}}\ and\ \bibinfo {author} {\bibfnamefont {L.}~\bibnamefont
  {Nordstr{\"o}m}},\ }\href@noop {} {\bibfield  {journal} {\bibinfo  {journal}
  {Phys. Rev. B}\ }\textbf {\bibinfo {volume} {66}},\ \bibinfo {pages} {014447}
  (\bibinfo {year} {2002})}\BibitemShut {NoStop}%
\bibitem [{\citenamefont {Balents}(2010)}]{Balents2010}%
  \BibitemOpen
  \bibfield  {author} {\bibinfo {author} {\bibfnamefont {L.}~\bibnamefont
  {Balents}},\ }\href@noop {} {\bibfield  {journal} {\bibinfo  {journal}
  {Nature}\ }\textbf {\bibinfo {volume} {464}},\ \bibinfo {pages} {199}
  (\bibinfo {year} {2010})}\BibitemShut {NoStop}%
\bibitem [{\citenamefont {Zhou}\ \emph {et~al.}(2017)\citenamefont {Zhou},
  \citenamefont {Kanoda},\ and\ \citenamefont {Ng}}]{Zhou2017}%
  \BibitemOpen
  \bibfield  {author} {\bibinfo {author} {\bibfnamefont {Y.}~\bibnamefont
  {Zhou}}, \bibinfo {author} {\bibfnamefont {K.}~\bibnamefont {Kanoda}}, \ and\
  \bibinfo {author} {\bibfnamefont {T.-K.}\ \bibnamefont {Ng}},\ }\href@noop {}
  {\bibfield  {journal} {\bibinfo  {journal} {Rev. Mod. Phys.}\ }\textbf
  {\bibinfo {volume} {89}},\ \bibinfo {pages} {025003} (\bibinfo {year}
  {2017})}\BibitemShut {NoStop}%
\bibitem [{\citenamefont {Yamanaka}\ \emph
  {et~al.}(2000{\natexlab{a}})\citenamefont {Yamanaka}, \citenamefont {Yamaki},
  \citenamefont {Shigeta}, \citenamefont {Nagao}, \citenamefont {Yoshioka},
  \citenamefont {Suzuki},\ and\ \citenamefont {Yamaguchi}}]{Yamanaka2000}%
  \BibitemOpen
  \bibfield  {author} {\bibinfo {author} {\bibfnamefont {S.}~\bibnamefont
  {Yamanaka}}, \bibinfo {author} {\bibfnamefont {D.}~\bibnamefont {Yamaki}},
  \bibinfo {author} {\bibfnamefont {Y.}~\bibnamefont {Shigeta}}, \bibinfo
  {author} {\bibfnamefont {H.}~\bibnamefont {Nagao}}, \bibinfo {author}
  {\bibfnamefont {Y.}~\bibnamefont {Yoshioka}}, \bibinfo {author}
  {\bibfnamefont {N.}~\bibnamefont {Suzuki}}, \ and\ \bibinfo {author}
  {\bibfnamefont {K.}~\bibnamefont {Yamaguchi}},\ }\href@noop {} {\bibfield
  {journal} {\bibinfo  {journal} {Int. J. Quant. Chem.}\ }\textbf {\bibinfo
  {volume} {80}},\ \bibinfo {pages} {664} (\bibinfo {year}
  {2000}{\natexlab{a}})}\BibitemShut {NoStop}%
\bibitem [{\citenamefont {Yamanaka}\ \emph
  {et~al.}(2000{\natexlab{b}})\citenamefont {Yamanaka}, \citenamefont {Yamaki},
  \citenamefont {Shigeta}, \citenamefont {Nagao},\ and\ \citenamefont
  {Yamaguchi}}]{Yamanaka2001}%
  \BibitemOpen
  \bibfield  {author} {\bibinfo {author} {\bibfnamefont {S.}~\bibnamefont
  {Yamanaka}}, \bibinfo {author} {\bibfnamefont {D.}~\bibnamefont {Yamaki}},
  \bibinfo {author} {\bibfnamefont {Y.}~\bibnamefont {Shigeta}}, \bibinfo
  {author} {\bibfnamefont {H.}~\bibnamefont {Nagao}}, \ and\ \bibinfo {author}
  {\bibfnamefont {K.}~\bibnamefont {Yamaguchi}},\ }\href@noop {} {\bibfield
  {journal} {\bibinfo  {journal} {Int. J. Quant. Chem.}\ }\textbf {\bibinfo
  {volume} {84}},\ \bibinfo {pages} {670} (\bibinfo {year}
  {2000}{\natexlab{b}})}\BibitemShut {NoStop}%
\bibitem [{\citenamefont {Freeman}\ and\ \citenamefont
  {Nakamura}(2004)}]{Freeman2004}%
  \BibitemOpen
  \bibfield  {author} {\bibinfo {author} {\bibfnamefont {A.~J.}\ \bibnamefont
  {Freeman}}\ and\ \bibinfo {author} {\bibfnamefont {K.}~\bibnamefont
  {Nakamura}},\ }\href@noop {} {\bibfield  {journal} {\bibinfo  {journal}
  {Phys. Stat. Sol. (b)}\ }\textbf {\bibinfo {volume} {241}},\ \bibinfo {pages}
  {1399} (\bibinfo {year} {2004})}\BibitemShut {NoStop}%
\bibitem [{\citenamefont {Peralta}\ \emph {et~al.}(2007)\citenamefont
  {Peralta}, \citenamefont {Scuseria},\ and\ \citenamefont
  {Frisch}}]{Peralta2007}%
  \BibitemOpen
  \bibfield  {author} {\bibinfo {author} {\bibfnamefont {J.~E.}\ \bibnamefont
  {Peralta}}, \bibinfo {author} {\bibfnamefont {G.~E.}\ \bibnamefont
  {Scuseria}}, \ and\ \bibinfo {author} {\bibfnamefont {M.~J.}\ \bibnamefont
  {Frisch}},\ }\href@noop {} {\bibfield  {journal} {\bibinfo  {journal} {Phys.
  Rev. B}\ }\textbf {\bibinfo {volume} {75}},\ \bibinfo {pages} {125119}
  (\bibinfo {year} {2007})}\BibitemShut {NoStop}%
\bibitem [{\citenamefont {Capelle}\ \emph {et~al.}(2001)\citenamefont
  {Capelle}, \citenamefont {Vignale},\ and\ \citenamefont
  {Gy{\"o}rffy}}]{Capelle2001}%
  \BibitemOpen
  \bibfield  {author} {\bibinfo {author} {\bibfnamefont {K.}~\bibnamefont
  {Capelle}}, \bibinfo {author} {\bibfnamefont {G.}~\bibnamefont {Vignale}}, \
  and\ \bibinfo {author} {\bibfnamefont {B.~L.}\ \bibnamefont {Gy{\"o}rffy}},\
  }\href@noop {} {\bibfield  {journal} {\bibinfo  {journal} {Phys. Rev. Lett.}\
  }\textbf {\bibinfo {volume} {87}},\ \bibinfo {pages} {206403} (\bibinfo
  {year} {2001})}\BibitemShut {NoStop}%
\bibitem [{\citenamefont {Kleinman}(1999)}]{Kleinman1999}%
  \BibitemOpen
  \bibfield  {author} {\bibinfo {author} {\bibfnamefont {L.}~\bibnamefont
  {Kleinman}},\ }\href@noop {} {\bibfield  {journal} {\bibinfo  {journal}
  {Phys. Rev. B}\ }\textbf {\bibinfo {volume} {59}},\ \bibinfo {pages} {3314}
  (\bibinfo {year} {1999})}\BibitemShut {NoStop}%
\bibitem [{\citenamefont {Katsnelson}\ and\ \citenamefont
  {Antropov}(2003)}]{Katsnelson2003}%
  \BibitemOpen
  \bibfield  {author} {\bibinfo {author} {\bibfnamefont {M.~I.}\ \bibnamefont
  {Katsnelson}}\ and\ \bibinfo {author} {\bibfnamefont {V.~P.}\ \bibnamefont
  {Antropov}},\ }\href@noop {} {\bibfield  {journal} {\bibinfo  {journal}
  {Phys. Rev. B}\ }\textbf {\bibinfo {volume} {67}},\ \bibinfo {pages} {140406}
  (\bibinfo {year} {2003})}\BibitemShut {NoStop}%
\bibitem [{\citenamefont {Sharma}\ \emph {et~al.}(2007)\citenamefont {Sharma},
  \citenamefont {Dewhurst}, \citenamefont {Ambrosch-Draxl}, \citenamefont
  {Kurth}, \citenamefont {Helbig}, \citenamefont {Pittalis}, \citenamefont
  {Shallcross}, \citenamefont {Nordstr{\"o}m},\ and\ \citenamefont
  {Gross}}]{Sharma2007}%
  \BibitemOpen
  \bibfield  {author} {\bibinfo {author} {\bibfnamefont {S.}~\bibnamefont
  {Sharma}}, \bibinfo {author} {\bibfnamefont {J.~K.}\ \bibnamefont
  {Dewhurst}}, \bibinfo {author} {\bibfnamefont {C.}~\bibnamefont
  {Ambrosch-Draxl}}, \bibinfo {author} {\bibfnamefont {S.}~\bibnamefont
  {Kurth}}, \bibinfo {author} {\bibfnamefont {N.}~\bibnamefont {Helbig}},
  \bibinfo {author} {\bibfnamefont {S.}~\bibnamefont {Pittalis}}, \bibinfo
  {author} {\bibfnamefont {S.}~\bibnamefont {Shallcross}}, \bibinfo {author}
  {\bibfnamefont {L.}~\bibnamefont {Nordstr{\"o}m}}, \ and\ \bibinfo {author}
  {\bibfnamefont {E.~K.~U.}\ \bibnamefont {Gross}},\ }\href@noop {} {\bibfield
  {journal} {\bibinfo  {journal} {Phys. Rev. Lett.}\ }\textbf {\bibinfo
  {volume} {98}},\ \bibinfo {pages} {196405} (\bibinfo {year}
  {2007})}\BibitemShut {NoStop}%
\bibitem [{\citenamefont {Scalmani}\ and\ \citenamefont
  {Frisch}(2012)}]{Scalmani2012}%
  \BibitemOpen
  \bibfield  {author} {\bibinfo {author} {\bibfnamefont {G.}~\bibnamefont
  {Scalmani}}\ and\ \bibinfo {author} {\bibfnamefont {M.~J.}\ \bibnamefont
  {Frisch}},\ }\href@noop {} {\bibfield  {journal} {\bibinfo  {journal} {J.
  Chem. Theor. Comput.}\ }\textbf {\bibinfo {volume} {8}},\ \bibinfo {pages}
  {2193} (\bibinfo {year} {2012})}\BibitemShut {NoStop}%
\bibitem [{\citenamefont {Bulik}\ \emph {et~al.}(2013)\citenamefont {Bulik},
  \citenamefont {Scalmani}, \citenamefont {Frisch},\ and\ \citenamefont
  {Scuseria}}]{Bulik2013}%
  \BibitemOpen
  \bibfield  {author} {\bibinfo {author} {\bibfnamefont {I.~W.}\ \bibnamefont
  {Bulik}}, \bibinfo {author} {\bibfnamefont {G.}~\bibnamefont {Scalmani}},
  \bibinfo {author} {\bibfnamefont {M.~J.}\ \bibnamefont {Frisch}}, \ and\
  \bibinfo {author} {\bibfnamefont {G.~E.}\ \bibnamefont {Scuseria}},\
  }\href@noop {} {\bibfield  {journal} {\bibinfo  {journal} {Phys. Rev. B}\
  }\textbf {\bibinfo {volume} {87}},\ \bibinfo {pages} {035117} (\bibinfo
  {year} {2013})}\BibitemShut {NoStop}%
\bibitem [{\citenamefont {Eich}\ and\ \citenamefont {Gross}(2013)}]{Eich2013a}%
  \BibitemOpen
  \bibfield  {author} {\bibinfo {author} {\bibfnamefont {F.~G.}\ \bibnamefont
  {Eich}}\ and\ \bibinfo {author} {\bibfnamefont {E.~K.~U.}\ \bibnamefont
  {Gross}},\ }\href@noop {} {\bibfield  {journal} {\bibinfo  {journal} {Phys.
  Rev. Lett.}\ }\textbf {\bibinfo {volume} {111}},\ \bibinfo {pages} {156401}
  (\bibinfo {year} {2013})}\BibitemShut {NoStop}%
\bibitem [{\citenamefont {Eich}\ \emph {et~al.}(2013)\citenamefont {Eich},
  \citenamefont {Pittalis},\ and\ \citenamefont {Vignale}}]{Eich2013b}%
  \BibitemOpen
  \bibfield  {author} {\bibinfo {author} {\bibfnamefont {F.~G.}\ \bibnamefont
  {Eich}}, \bibinfo {author} {\bibfnamefont {S.}~\bibnamefont {Pittalis}}, \
  and\ \bibinfo {author} {\bibfnamefont {G.}~\bibnamefont {Vignale}},\
  }\href@noop {} {\bibfield  {journal} {\bibinfo  {journal} {Phys. Rev. B}\
  }\textbf {\bibinfo {volume} {88}},\ \bibinfo {pages} {245102} (\bibinfo
  {year} {2013})}\BibitemShut {NoStop}%
\bibitem [{\citenamefont {Pittalis}\ \emph {et~al.}(2017)\citenamefont
  {Pittalis}, \citenamefont {Vignale},\ and\ \citenamefont
  {Eich}}]{Pittalis2017}%
  \BibitemOpen
  \bibfield  {author} {\bibinfo {author} {\bibfnamefont {S.}~\bibnamefont
  {Pittalis}}, \bibinfo {author} {\bibfnamefont {G.}~\bibnamefont {Vignale}}, \
  and\ \bibinfo {author} {\bibfnamefont {F.~G.}\ \bibnamefont {Eich}},\
  }\href@noop {} {\bibfield  {journal} {\bibinfo  {journal} {Phys. Rev. B}\
  }\textbf {\bibinfo {volume} {96}},\ \bibinfo {pages} {035141} (\bibinfo
  {year} {2017})}\BibitemShut {NoStop}%
\bibitem [{\citenamefont {Goings}\ \emph {et~al.}(2018)\citenamefont {Goings},
  \citenamefont {Egidi},\ and\ \citenamefont {Li}}]{Goings2018}%
  \BibitemOpen
  \bibfield  {author} {\bibinfo {author} {\bibfnamefont {J.~J.}\ \bibnamefont
  {Goings}}, \bibinfo {author} {\bibfnamefont {F.}~\bibnamefont {Egidi}}, \
  and\ \bibinfo {author} {\bibfnamefont {X.}~\bibnamefont {Li}},\ }\href@noop
  {} {\bibfield  {journal} {\bibinfo  {journal} {Int. J. Quantum Chem.}\
  }\textbf {\bibinfo {volume} {118}},\ \bibinfo {pages} {e25398} (\bibinfo
  {year} {2018})}\BibitemShut {NoStop}%
\bibitem [{\citenamefont {Singwi}\ \emph {et~al.}(1968)\citenamefont {Singwi},
  \citenamefont {Sj{\"o}lander}, \citenamefont {Tosi},\ and\ \citenamefont
  {Land}}]{Singwi1968}%
  \BibitemOpen
  \bibfield  {author} {\bibinfo {author} {\bibfnamefont {K.~S.}\ \bibnamefont
  {Singwi}}, \bibinfo {author} {\bibfnamefont {A.}~\bibnamefont
  {Sj{\"o}lander}}, \bibinfo {author} {\bibfnamefont {M.~P.}\ \bibnamefont
  {Tosi}}, \ and\ \bibinfo {author} {\bibfnamefont {R.~H.}\ \bibnamefont
  {Land}},\ }\href@noop {} {\bibfield  {journal} {\bibinfo  {journal} {Phys.
  Rev.}\ }\textbf {\bibinfo {volume} {176}},\ \bibinfo {pages} {589} (\bibinfo
  {year} {1968})}\BibitemShut {NoStop}%
\bibitem [{\citenamefont {Singwi}\ and\ \citenamefont
  {Tosi}(1989)}]{Singwi1981}%
  \BibitemOpen
  \bibfield  {author} {\bibinfo {author} {\bibfnamefont {K.~S.}\ \bibnamefont
  {Singwi}}\ and\ \bibinfo {author} {\bibfnamefont {M.~P.}\ \bibnamefont
  {Tosi}},\ }\href@noop {} {\bibfield  {journal} {\bibinfo  {journal} {Solid
  State Phys.}\ }\textbf {\bibinfo {volume} {36}},\ \bibinfo {pages} {177}
  (\bibinfo {year} {1989})}\BibitemShut {NoStop}%
\bibitem [{\citenamefont {Giuliani}\ and\ \citenamefont
  {Vignale}(2005)}]{GiulianiVignale}%
  \BibitemOpen
  \bibfield  {author} {\bibinfo {author} {\bibfnamefont {G.~F.}\ \bibnamefont
  {Giuliani}}\ and\ \bibinfo {author} {\bibfnamefont {G.}~\bibnamefont
  {Vignale}},\ }\href@noop {} {\emph {\bibinfo {title} {Quantum Theory of the
  Electron Liquid}}}\ (\bibinfo  {publisher} {Cambridge University Press},\
  \bibinfo {address} {Cambridge},\ \bibinfo {year} {2005})\BibitemShut
  {NoStop}%
\bibitem [{\citenamefont {Lobo}\ \emph {et~al.}(1969)\citenamefont {Lobo},
  \citenamefont {Singwi},\ and\ \citenamefont {Tosi}}]{Lobo1969}%
  \BibitemOpen
  \bibfield  {author} {\bibinfo {author} {\bibfnamefont {R.}~\bibnamefont
  {Lobo}}, \bibinfo {author} {\bibfnamefont {K.~S.}\ \bibnamefont {Singwi}}, \
  and\ \bibinfo {author} {\bibfnamefont {M.~P.}\ \bibnamefont {Tosi}},\
  }\href@noop {} {\bibfield  {journal} {\bibinfo  {journal} {Phys. Rev.}\
  }\textbf {\bibinfo {volume} {186}},\ \bibinfo {pages} {470} (\bibinfo {year}
  {1969})}\BibitemShut {NoStop}%
\bibitem [{\citenamefont {Vashishta}\ and\ \citenamefont
  {Singwi}(1972)}]{Vashishta1972}%
  \BibitemOpen
  \bibfield  {author} {\bibinfo {author} {\bibfnamefont {P.}~\bibnamefont
  {Vashishta}}\ and\ \bibinfo {author} {\bibfnamefont {K.~S.}\ \bibnamefont
  {Singwi}},\ }\href@noop {} {\bibfield  {journal} {\bibinfo  {journal} {Phys.
  Rev. B}\ }\textbf {\bibinfo {volume} {6}},\ \bibinfo {pages} {875} (\bibinfo
  {year} {1972})}\BibitemShut {NoStop}%
\bibitem [{\citenamefont {Holas}\ and\ \citenamefont
  {Rahman}(1987)}]{Holas1987}%
  \BibitemOpen
  \bibfield  {author} {\bibinfo {author} {\bibfnamefont {A.}~\bibnamefont
  {Holas}}\ and\ \bibinfo {author} {\bibfnamefont {S.}~\bibnamefont {Rahman}},\
  }\href@noop {} {\bibfield  {journal} {\bibinfo  {journal} {Phys. Rev. B}\
  }\textbf {\bibinfo {volume} {35}},\ \bibinfo {pages} {2720} (\bibinfo {year}
  {1987})}\BibitemShut {NoStop}%
\bibitem [{\citenamefont {Moudgil}\ \emph
  {et~al.}(1995{\natexlab{a}})\citenamefont {Moudgil}, \citenamefont
  {Ahluwalia},\ and\ \citenamefont {Pathak}}]{Moudgil1995a}%
  \BibitemOpen
  \bibfield  {author} {\bibinfo {author} {\bibfnamefont {R.~K.}\ \bibnamefont
  {Moudgil}}, \bibinfo {author} {\bibfnamefont {P.~K.}\ \bibnamefont
  {Ahluwalia}}, \ and\ \bibinfo {author} {\bibfnamefont {K.~N.}\ \bibnamefont
  {Pathak}},\ }\href@noop {} {\bibfield  {journal} {\bibinfo  {journal} {Phys.
  Rev. B}\ }\textbf {\bibinfo {volume} {51}},\ \bibinfo {pages} {1575}
  (\bibinfo {year} {1995}{\natexlab{a}})}\BibitemShut {NoStop}%
\bibitem [{\citenamefont {Moudgil}\ \emph
  {et~al.}(1995{\natexlab{b}})\citenamefont {Moudgil}, \citenamefont
  {Ahluwalia},\ and\ \citenamefont {Pathak}}]{Moudgil1995b}%
  \BibitemOpen
  \bibfield  {author} {\bibinfo {author} {\bibfnamefont {R.~K.}\ \bibnamefont
  {Moudgil}}, \bibinfo {author} {\bibfnamefont {P.~K.}\ \bibnamefont
  {Ahluwalia}}, \ and\ \bibinfo {author} {\bibfnamefont {K.~N.}\ \bibnamefont
  {Pathak}},\ }\href@noop {} {\bibfield  {journal} {\bibinfo  {journal} {Phys.
  Rev. B}\ }\textbf {\bibinfo {volume} {52}},\ \bibinfo {pages} {11945}
  (\bibinfo {year} {1995}{\natexlab{b}})}\BibitemShut {NoStop}%
\bibitem [{\citenamefont {Calmels}\ and\ \citenamefont
  {Gold}(1995)}]{Calmels1995}%
  \BibitemOpen
  \bibfield  {author} {\bibinfo {author} {\bibfnamefont {L.}~\bibnamefont
  {Calmels}}\ and\ \bibinfo {author} {\bibfnamefont {A.}~\bibnamefont {Gold}},\
  }\href@noop {} {\bibfield  {journal} {\bibinfo  {journal} {Phys. Rev. B}\
  }\textbf {\bibinfo {volume} {52}},\ \bibinfo {pages} {10841} (\bibinfo {year}
  {1995})}\BibitemShut {NoStop}%
\bibitem [{\citenamefont {Calmels}\ and\ \citenamefont
  {Gold}(1997{\natexlab{a}})}]{Calmels1997a}%
  \BibitemOpen
  \bibfield  {author} {\bibinfo {author} {\bibfnamefont {L.}~\bibnamefont
  {Calmels}}\ and\ \bibinfo {author} {\bibfnamefont {A.}~\bibnamefont {Gold}},\
  }\href@noop {} {\bibfield  {journal} {\bibinfo  {journal} {Phys. Rev. B}\
  }\textbf {\bibinfo {volume} {56}},\ \bibinfo {pages} {1762} (\bibinfo {year}
  {1997}{\natexlab{a}})}\BibitemShut {NoStop}%
\bibitem [{\citenamefont {Calmels}\ and\ \citenamefont
  {Gold}(1997{\natexlab{b}})}]{Calmels1997b}%
  \BibitemOpen
  \bibfield  {author} {\bibinfo {author} {\bibfnamefont {L.}~\bibnamefont
  {Calmels}}\ and\ \bibinfo {author} {\bibfnamefont {A.}~\bibnamefont {Gold}},\
  }\href@noop {} {\bibfield  {journal} {\bibinfo  {journal} {Europhys. Lett.}\
  }\textbf {\bibinfo {volume} {39}},\ \bibinfo {pages} {539} (\bibinfo {year}
  {1997}{\natexlab{b}})}\BibitemShut {NoStop}%
\bibitem [{\citenamefont {Kim}\ \emph {et~al.}(2001)\citenamefont {Kim},
  \citenamefont {Yi},\ and\ \citenamefont {Choi}}]{Kim2001}%
  \BibitemOpen
  \bibfield  {author} {\bibinfo {author} {\bibfnamefont {J.}~\bibnamefont
  {Kim}}, \bibinfo {author} {\bibfnamefont {K.~S.}\ \bibnamefont {Yi}}, \ and\
  \bibinfo {author} {\bibfnamefont {S.~D.}\ \bibnamefont {Choi}},\ }\href@noop
  {} {\bibfield  {journal} {\bibinfo  {journal} {J. Korean Phys. Soc.}\
  }\textbf {\bibinfo {volume} {38}},\ \bibinfo {pages} {729} (\bibinfo {year}
  {2001})}\BibitemShut {NoStop}%
\bibitem [{\citenamefont {Ta{\c{s}}}\ and\ \citenamefont
  {Tomak}(2004)}]{Tas2004}%
  \BibitemOpen
  \bibfield  {author} {\bibinfo {author} {\bibfnamefont {M.}~\bibnamefont
  {Ta{\c{s}}}}\ and\ \bibinfo {author} {\bibfnamefont {M.}~\bibnamefont
  {Tomak}},\ }\href@noop {} {\bibfield  {journal} {\bibinfo  {journal} {Phys.
  Rev. B}\ }\textbf {\bibinfo {volume} {70}},\ \bibinfo {pages} {235305}
  (\bibinfo {year} {2004})}\BibitemShut {NoStop}%
\bibitem [{\citenamefont {Dobson}\ \emph {et~al.}(2004)\citenamefont {Dobson},
  \citenamefont {Le},\ and\ \citenamefont {Vignale}}]{Dobson2004}%
  \BibitemOpen
  \bibfield  {author} {\bibinfo {author} {\bibfnamefont {J.~F.}\ \bibnamefont
  {Dobson}}, \bibinfo {author} {\bibfnamefont {H.~M.}\ \bibnamefont {Le}}, \
  and\ \bibinfo {author} {\bibfnamefont {G.}~\bibnamefont {Vignale}},\
  }\href@noop {} {\bibfield  {journal} {\bibinfo  {journal} {Phys. Rev. B}\
  }\textbf {\bibinfo {volume} {70}},\ \bibinfo {pages} {205126} (\bibinfo
  {year} {2004})}\BibitemShut {NoStop}%
\bibitem [{\citenamefont {Kumar}\ \emph {et~al.}(2009)\citenamefont {Kumar},
  \citenamefont {Garg},\ and\ \citenamefont {Moudgil}}]{Kumar2009}%
  \BibitemOpen
  \bibfield  {author} {\bibinfo {author} {\bibfnamefont {K.}~\bibnamefont
  {Kumar}}, \bibinfo {author} {\bibfnamefont {V.}~\bibnamefont {Garg}}, \ and\
  \bibinfo {author} {\bibfnamefont {R.~K.}\ \bibnamefont {Moudgil}},\
  }\href@noop {} {\bibfield  {journal} {\bibinfo  {journal} {Phys. Rev. B}\
  }\textbf {\bibinfo {volume} {79}},\ \bibinfo {pages} {115304} (\bibinfo
  {year} {2009})}\BibitemShut {NoStop}%
\bibitem [{\citenamefont {Yoshizawa}\ and\ \citenamefont
  {Takada}(2009)}]{Yoshizawa2009}%
  \BibitemOpen
  \bibfield  {author} {\bibinfo {author} {\bibfnamefont {K.}~\bibnamefont
  {Yoshizawa}}\ and\ \bibinfo {author} {\bibfnamefont {Y.}~\bibnamefont
  {Takada}},\ }\href@noop {} {\bibfield  {journal} {\bibinfo  {journal} {J.
  Phys.: Condens. Matter}\ }\textbf {\bibinfo {volume} {21}},\ \bibinfo {pages}
  {064104} (\bibinfo {year} {2009})}\BibitemShut {NoStop}%
\bibitem [{\citenamefont {Hedayati}\ and\ \citenamefont
  {Vignale}(1989)}]{Hedayati1989}%
  \BibitemOpen
  \bibfield  {author} {\bibinfo {author} {\bibfnamefont {M.~R.}\ \bibnamefont
  {Hedayati}}\ and\ \bibinfo {author} {\bibfnamefont {G.}~\bibnamefont
  {Vignale}},\ }\href@noop {} {\bibfield  {journal} {\bibinfo  {journal} {Phys.
  Rev. B}\ }\textbf {\bibinfo {volume} {40}},\ \bibinfo {pages} {9044}
  (\bibinfo {year} {1989})}\BibitemShut {NoStop}%
\bibitem [{\citenamefont {Dobson}\ \emph {et~al.}(2002)\citenamefont {Dobson},
  \citenamefont {Wang},\ and\ \citenamefont {Gould}}]{Dobson2002}%
  \BibitemOpen
  \bibfield  {author} {\bibinfo {author} {\bibfnamefont {J.~F.}\ \bibnamefont
  {Dobson}}, \bibinfo {author} {\bibfnamefont {J.}~\bibnamefont {Wang}}, \ and\
  \bibinfo {author} {\bibfnamefont {T.}~\bibnamefont {Gould}},\ }\href@noop {}
  {\bibfield  {journal} {\bibinfo  {journal} {Phys. Rev. B}\ }\textbf {\bibinfo
  {volume} {66}},\ \bibinfo {pages} {081108} (\bibinfo {year}
  {2002})}\BibitemShut {NoStop}%
\bibitem [{\citenamefont {Dobson}(2009)}]{Dobson2009}%
  \BibitemOpen
  \bibfield  {author} {\bibinfo {author} {\bibfnamefont {J.~F.}\ \bibnamefont
  {Dobson}},\ }\href@noop {} {\bibfield  {journal} {\bibinfo  {journal} {Phys.
  Chem. Chem. Phys.}\ }\textbf {\bibinfo {volume} {11}},\ \bibinfo {pages}
  {4528} (\bibinfo {year} {2009})}\BibitemShut {NoStop}%
\bibitem [{\citenamefont {Gould}\ and\ \citenamefont
  {Dobson}(2012)}]{Gould2012}%
  \BibitemOpen
  \bibfield  {author} {\bibinfo {author} {\bibfnamefont {T.}~\bibnamefont
  {Gould}}\ and\ \bibinfo {author} {\bibfnamefont {J.~F.}\ \bibnamefont
  {Dobson}},\ }\href@noop {} {\bibfield  {journal} {\bibinfo  {journal} {Phys.
  Rev. A}\ }\textbf {\bibinfo {volume} {85}},\ \bibinfo {pages} {062504}
  (\bibinfo {year} {2012})}\BibitemShut {NoStop}%
\bibitem [{\citenamefont {Slater}(1951)}]{Slater1951}%
  \BibitemOpen
  \bibfield  {author} {\bibinfo {author} {\bibfnamefont {J.~C.}\ \bibnamefont
  {Slater}},\ }\href@noop {} {\bibfield  {journal} {\bibinfo  {journal} {Phys.
  Rev.}\ }\textbf {\bibinfo {volume} {81}},\ \bibinfo {pages} {385} (\bibinfo
  {year} {1951})}\BibitemShut {NoStop}%
\bibitem [{\citenamefont {Talman}\ and\ \citenamefont
  {Shadwick}(1976)}]{Talman1976}%
  \BibitemOpen
  \bibfield  {author} {\bibinfo {author} {\bibfnamefont {J.~D.}\ \bibnamefont
  {Talman}}\ and\ \bibinfo {author} {\bibfnamefont {W.~F.}\ \bibnamefont
  {Shadwick}},\ }\href@noop {} {\bibfield  {journal} {\bibinfo  {journal}
  {Phys. Rev. A}\ }\textbf {\bibinfo {volume} {14}},\ \bibinfo {pages} {36}
  (\bibinfo {year} {1976})}\BibitemShut {NoStop}%
\bibitem [{\citenamefont {Krieger}\ \emph {et~al.}(1992)\citenamefont
  {Krieger}, \citenamefont {Li},\ and\ \citenamefont {Iafrate}}]{Krieger1992}%
  \BibitemOpen
  \bibfield  {author} {\bibinfo {author} {\bibfnamefont {J.~B.}\ \bibnamefont
  {Krieger}}, \bibinfo {author} {\bibfnamefont {Y.}~\bibnamefont {Li}}, \ and\
  \bibinfo {author} {\bibfnamefont {G.~J.}\ \bibnamefont {Iafrate}},\
  }\href@noop {} {\bibfield  {journal} {\bibinfo  {journal} {Phys. Rev. A}\
  }\textbf {\bibinfo {volume} {45}},\ \bibinfo {pages} {101} (\bibinfo {year}
  {1992})}\BibitemShut {NoStop}%
\bibitem [{\citenamefont {K{\"u}mmel}\ and\ \citenamefont
  {Kronik}(2008)}]{Kummel2008}%
  \BibitemOpen
  \bibfield  {author} {\bibinfo {author} {\bibfnamefont {S.}~\bibnamefont
  {K{\"u}mmel}}\ and\ \bibinfo {author} {\bibfnamefont {L.}~\bibnamefont
  {Kronik}},\ }\href@noop {} {\bibfield  {journal} {\bibinfo  {journal} {Rev.
  Mod. Phys.}\ }\textbf {\bibinfo {volume} {80}},\ \bibinfo {pages} {3}
  (\bibinfo {year} {2008})}\BibitemShut {NoStop}%
\bibitem [{\citenamefont {Carrascal}\ \emph {et~al.}(2015)\citenamefont
  {Carrascal}, \citenamefont {Ferrer}, \citenamefont {Smith},\ and\
  \citenamefont {Burke}}]{Carrascal2015}%
  \BibitemOpen
  \bibfield  {author} {\bibinfo {author} {\bibfnamefont {D.~J.}\ \bibnamefont
  {Carrascal}}, \bibinfo {author} {\bibfnamefont {J.}~\bibnamefont {Ferrer}},
  \bibinfo {author} {\bibfnamefont {J.~C.}\ \bibnamefont {Smith}}, \ and\
  \bibinfo {author} {\bibfnamefont {K.}~\bibnamefont {Burke}},\ }\href@noop {}
  {\bibfield  {journal} {\bibinfo  {journal} {J. Phys.: Condens. Matter}\
  }\textbf {\bibinfo {volume} {27}},\ \bibinfo {pages} {393001} (\bibinfo
  {year} {2015})}\BibitemShut {NoStop}%
\bibitem [{\citenamefont {Carrascal}\ \emph {et~al.}(2018)\citenamefont
  {Carrascal}, \citenamefont {Ferrer}, \citenamefont {Maitra},\ and\
  \citenamefont {Burke}}]{Carrascal2018}%
  \BibitemOpen
  \bibfield  {author} {\bibinfo {author} {\bibfnamefont {D.~J.}\ \bibnamefont
  {Carrascal}}, \bibinfo {author} {\bibfnamefont {J.}~\bibnamefont {Ferrer}},
  \bibinfo {author} {\bibfnamefont {N.~T.}\ \bibnamefont {Maitra}}, \ and\
  \bibinfo {author} {\bibfnamefont {K.}~\bibnamefont {Burke}},\ }\href@noop {}
  {\bibfield  {journal} {\bibinfo  {journal} {arXiv:1802.09988}\ } (\bibinfo
  {year} {2018})}\BibitemShut {NoStop}%
\bibitem [{\citenamefont {Balcerzak}\ and\ \citenamefont
  {Sza{\l}owski}(2017)}]{Balcerzak2017}%
  \BibitemOpen
  \bibfield  {author} {\bibinfo {author} {\bibfnamefont {T.}~\bibnamefont
  {Balcerzak}}\ and\ \bibinfo {author} {\bibfnamefont {K.}~\bibnamefont
  {Sza{\l}owski}},\ }\href@noop {} {\bibfield  {journal} {\bibinfo  {journal}
  {Physica A}\ }\textbf {\bibinfo {volume} {468}},\ \bibinfo {pages} {252}
  (\bibinfo {year} {2017})}\BibitemShut {NoStop}%
\bibitem [{\citenamefont {Balcerzak}\ and\ \citenamefont
  {Sza{\l}owski}(2018)}]{Balcerzak2018}%
  \BibitemOpen
  \bibfield  {author} {\bibinfo {author} {\bibfnamefont {T.}~\bibnamefont
  {Balcerzak}}\ and\ \bibinfo {author} {\bibfnamefont {K.}~\bibnamefont
  {Sza{\l}owski}},\ }\href@noop {} {\bibfield  {journal} {\bibinfo  {journal}
  {Physica A}\ }\textbf {\bibinfo {volume} {499}},\ \bibinfo {pages} {395}
  (\bibinfo {year} {2018})}\BibitemShut {NoStop}%
\bibitem [{\citenamefont {Vignale}\ and\ \citenamefont
  {Rasolt}(1987)}]{Vignale1987}%
  \BibitemOpen
  \bibfield  {author} {\bibinfo {author} {\bibfnamefont {G.}~\bibnamefont
  {Vignale}}\ and\ \bibinfo {author} {\bibfnamefont {M.}~\bibnamefont
  {Rasolt}},\ }\href@noop {} {\bibfield  {journal} {\bibinfo  {journal} {Phys.
  Rev. Lett.}\ }\textbf {\bibinfo {volume} {59}},\ \bibinfo {pages} {2360}
  (\bibinfo {year} {1987})},\ \bibinfo {note} {erratum: {\em ibid.} {\bf 62},
  115 (1989).}\BibitemShut {Stop}%
\bibitem [{\citenamefont {Vignale}\ and\ \citenamefont
  {Rasolt}(1988)}]{Vignale1988}%
  \BibitemOpen
  \bibfield  {author} {\bibinfo {author} {\bibfnamefont {G.}~\bibnamefont
  {Vignale}}\ and\ \bibinfo {author} {\bibfnamefont {M.}~\bibnamefont
  {Rasolt}},\ }\href@noop {} {\bibfield  {journal} {\bibinfo  {journal} {Phys.
  Rev. B}\ }\textbf {\bibinfo {volume} {37}},\ \bibinfo {pages} {10685}
  (\bibinfo {year} {1988})},\ \bibinfo {note} {erratum: {\em ibid.} {\bf 39},
  5475 (1989).}\BibitemShut {Stop}%
\bibitem [{\citenamefont {von Barth}\ and\ \citenamefont
  {Hedin}(1972)}]{Barth1972}%
  \BibitemOpen
  \bibfield  {author} {\bibinfo {author} {\bibfnamefont {U.}~\bibnamefont {von
  Barth}}\ and\ \bibinfo {author} {\bibfnamefont {L.}~\bibnamefont {Hedin}},\
  }\href@noop {} {\bibfield  {journal} {\bibinfo  {journal} {J. Phys. C}\
  }\textbf {\bibinfo {volume} {5}},\ \bibinfo {pages} {1629} (\bibinfo {year}
  {1972})}\BibitemShut {NoStop}%
\bibitem [{\citenamefont {Gunnarsson}\ and\ \citenamefont
  {Lundqvist}(1976)}]{Gunnarsson1976}%
  \BibitemOpen
  \bibfield  {author} {\bibinfo {author} {\bibfnamefont {O.}~\bibnamefont
  {Gunnarsson}}\ and\ \bibinfo {author} {\bibfnamefont {B.~I.}\ \bibnamefont
  {Lundqvist}},\ }\href@noop {} {\bibfield  {journal} {\bibinfo  {journal}
  {Phys. Rev. B}\ }\textbf {\bibinfo {volume} {13}},\ \bibinfo {pages} {4274}
  (\bibinfo {year} {1976})}\BibitemShut {NoStop}%
\bibitem [{sup()}]{supp}%
  \BibitemOpen
  \href@noop {} {}\bibinfo {note} {See Supplemental Material at http://... for
  the following details: (1) Derivation of the OEP, KLI and Slater potentials.
  (2) Derivation of the PGG kernel. (3) Derivation of the spectral
  representation of the full response function. (4) Derivation of the
  coupling-constant expression for the xc energy.}\BibitemShut {Stop}%
\bibitem [{\citenamefont {Sharp}\ and\ \citenamefont
  {Horton}(1953)}]{Sharp1953}%
  \BibitemOpen
  \bibfield  {author} {\bibinfo {author} {\bibfnamefont {R.~T.}\ \bibnamefont
  {Sharp}}\ and\ \bibinfo {author} {\bibfnamefont {G.~K.}\ \bibnamefont
  {Horton}},\ }\href@noop {} {\bibfield  {journal} {\bibinfo  {journal} {Phys.
  Rev.}\ }\textbf {\bibinfo {volume} {30}},\ \bibinfo {pages} {317} (\bibinfo
  {year} {1953})}\BibitemShut {NoStop}%
\bibitem [{\citenamefont {Engel}\ and\ \citenamefont
  {Dreizler}(2011)}]{EngelDreizler}%
  \BibitemOpen
  \bibfield  {author} {\bibinfo {author} {\bibfnamefont {E.}~\bibnamefont
  {Engel}}\ and\ \bibinfo {author} {\bibfnamefont {R.~M.}\ \bibnamefont
  {Dreizler}},\ }\href@noop {} {\emph {\bibinfo {title} {Density Functional
  Theory: an advanced course}}}\ (\bibinfo  {publisher} {Springer},\ \bibinfo
  {address} {Berlin},\ \bibinfo {year} {2011})\BibitemShut {NoStop}%
\bibitem [{\citenamefont {Petersilka}\ \emph {et~al.}(1996)\citenamefont
  {Petersilka}, \citenamefont {Gossmann},\ and\ \citenamefont
  {Gross}}]{Petersilka1996}%
  \BibitemOpen
  \bibfield  {author} {\bibinfo {author} {\bibfnamefont {M.}~\bibnamefont
  {Petersilka}}, \bibinfo {author} {\bibfnamefont {U.~J.}\ \bibnamefont
  {Gossmann}}, \ and\ \bibinfo {author} {\bibfnamefont {E.~K.~U.}\ \bibnamefont
  {Gross}},\ }\href@noop {} {\bibfield  {journal} {\bibinfo  {journal} {Phys.
  Rev. Lett.}\ }\textbf {\bibinfo {volume} {76}},\ \bibinfo {pages} {1212}
  (\bibinfo {year} {1996})}\BibitemShut {NoStop}%
\bibitem [{\citenamefont {Petersilka}\ \emph {et~al.}(1998)\citenamefont
  {Petersilka}, \citenamefont {Gossmann},\ and\ \citenamefont
  {Gross}}]{Petersilka1998}%
  \BibitemOpen
  \bibfield  {author} {\bibinfo {author} {\bibfnamefont {M.}~\bibnamefont
  {Petersilka}}, \bibinfo {author} {\bibfnamefont {U.~J.}\ \bibnamefont
  {Gossmann}}, \ and\ \bibinfo {author} {\bibfnamefont {E.~K.~U.}\ \bibnamefont
  {Gross}},\ }in\ \href@noop {} {\emph {\bibinfo {booktitle} {Electronic
  density functional theory: recent progress and new directions}}},\ \bibinfo
  {editor} {edited by\ \bibinfo {editor} {\bibfnamefont {J.~F.}\ \bibnamefont
  {Dobson}}, \bibinfo {editor} {\bibfnamefont {G.}~\bibnamefont {Vignale}}, \
  and\ \bibinfo {editor} {\bibfnamefont {M.~P.}\ \bibnamefont {Das}}}\
  (\bibinfo  {publisher} {Plenum},\ \bibinfo {address} {New York},\ \bibinfo
  {year} {1998})\ pp.\ \bibinfo {pages} {177--97}\BibitemShut {NoStop}%
\bibitem [{\citenamefont {Ullrich}(2012)}]{Ullrich2012}%
  \BibitemOpen
  \bibfield  {author} {\bibinfo {author} {\bibfnamefont {C.~A.}\ \bibnamefont
  {Ullrich}},\ }\href@noop {} {\emph {\bibinfo {title} {Time-dependent
  density-functional theory: concepts and applications}}}\ (\bibinfo
  {publisher} {Oxford University Press},\ \bibinfo {address} {Oxford},\
  \bibinfo {year} {2012})\BibitemShut {NoStop}%
\bibitem [{\citenamefont {Lein}\ and\ \citenamefont {Gross}(2006)}]{Lein2006}%
  \BibitemOpen
  \bibfield  {author} {\bibinfo {author} {\bibfnamefont {M.}~\bibnamefont
  {Lein}}\ and\ \bibinfo {author} {\bibfnamefont {E.~K.~U.}\ \bibnamefont
  {Gross}},\ }in\ \href@noop {} {\emph {\bibinfo {booktitle} {Time-Dependent
  Density Functional Theory}}},\ \bibinfo {series} {Lecture Notes in Physics},
  Vol.\ \bibinfo {volume} {706},\ \bibinfo {editor} {edited by\ \bibinfo
  {editor} {\bibfnamefont {M.~A.~L.}\ \bibnamefont {Marques}}, \bibinfo
  {editor} {\bibfnamefont {C.~A.}\ \bibnamefont {Ullrich}}, \bibinfo {editor}
  {\bibfnamefont {F.}~\bibnamefont {Nogueira}}, \bibinfo {editor}
  {\bibfnamefont {A.}~\bibnamefont {Rubio}}, \bibinfo {editor} {\bibfnamefont
  {K.}~\bibnamefont {Burke}}, \ and\ \bibinfo {editor} {\bibfnamefont
  {E.~K.~U.}\ \bibnamefont {Gross}}}\ (\bibinfo  {publisher} {Springer},\
  \bibinfo {address} {Berlin},\ \bibinfo {year} {2006})\ pp.\ \bibinfo {pages}
  {423--34}\BibitemShut {NoStop}%
\bibitem [{\citenamefont {Furche}(2001)}]{Furche2001}%
  \BibitemOpen
  \bibfield  {author} {\bibinfo {author} {\bibfnamefont {F.}~\bibnamefont
  {Furche}},\ }\href@noop {} {\bibfield  {journal} {\bibinfo  {journal} {Phys.
  Rev. B}\ }\textbf {\bibinfo {volume} {64}},\ \bibinfo {pages} {195120}
  (\bibinfo {year} {2001})}\BibitemShut {NoStop}%
\bibitem [{\citenamefont {Casida}(1995)}]{Casida1995}%
  \BibitemOpen
  \bibfield  {author} {\bibinfo {author} {\bibfnamefont {M.~E.}\ \bibnamefont
  {Casida}},\ }in\ \href@noop {} {\emph {\bibinfo {booktitle} {Recent Advances
  in Density Functional Methods}}},\ \bibinfo {series} {Recent Advances in
  Computational Chemistry}, Vol.~\bibinfo {volume} {1},\ \bibinfo {editor}
  {edited by\ \bibinfo {editor} {\bibfnamefont {D.~E.}\ \bibnamefont {Chong}}}\
  (\bibinfo  {publisher} {World Scientific},\ \bibinfo {address} {Singapore},\
  \bibinfo {year} {1995})\ pp.\ \bibinfo {pages} {155--92}\BibitemShut
  {NoStop}%
\bibitem [{\citenamefont {Fran{\c{c}}a}\ \emph {et~al.}(2012)\citenamefont
  {Fran{\c{c}}a}, \citenamefont {Vieira},\ and\ \citenamefont
  {Capelle}}]{Franca2012}%
  \BibitemOpen
  \bibfield  {author} {\bibinfo {author} {\bibfnamefont {V.~V.}\ \bibnamefont
  {Fran{\c{c}}a}}, \bibinfo {author} {\bibfnamefont {D.}~\bibnamefont
  {Vieira}}, \ and\ \bibinfo {author} {\bibfnamefont {K.}~\bibnamefont
  {Capelle}},\ }\href@noop {} {\bibfield  {journal} {\bibinfo  {journal} {New
  J. Phys.}\ }\textbf {\bibinfo {volume} {14}},\ \bibinfo {pages} {073021}
  (\bibinfo {year} {2012})}\BibitemShut {NoStop}%
\bibitem [{\citenamefont {Capelle}\ and\ \citenamefont {{Campo
  Jr.}}(2013)}]{Capelle2013}%
  \BibitemOpen
  \bibfield  {author} {\bibinfo {author} {\bibfnamefont {K.}~\bibnamefont
  {Capelle}}\ and\ \bibinfo {author} {\bibfnamefont {V.~L.}\ \bibnamefont
  {{Campo Jr.}}},\ }\href@noop {} {\bibfield  {journal} {\bibinfo  {journal}
  {Phys. Rep.}\ }\textbf {\bibinfo {volume} {528}},\ \bibinfo {pages} {91}
  (\bibinfo {year} {2013})}\BibitemShut {NoStop}%
\bibitem [{\citenamefont {Ullrich}(2005)}]{Ullrich2005}%
  \BibitemOpen
  \bibfield  {author} {\bibinfo {author} {\bibfnamefont {C.~A.}\ \bibnamefont
  {Ullrich}},\ }\href@noop {} {\bibfield  {journal} {\bibinfo  {journal} {Phys.
  Rev. B}\ }\textbf {\bibinfo {volume} {72}},\ \bibinfo {pages} {073102}
  (\bibinfo {year} {2005})}\BibitemShut {NoStop}%
\end{thebibliography}%

\end{document}